\documentclass[prd,aps,nofootinbib,superscriptaddress]{revtex4}
\usepackage{epsfig}
\usepackage{amsmath, slashed}
\usepackage{xcolor}
\usepackage{appendix}

\usepackage[normalem]{ulem}

\newcommand{\uvec}{\boldsymbol}
\newcommand{\ud}{\mathrm{d}}
\newcommand{\ucal}[1]{\mathcal{#1}}

\usepackage{bm}
\usepackage{amsfonts}
\definecolor{myGreen}{rgb}{0.2,0.72,0.2}

\renewcommand{\[}{\begin{equation}}
\renewcommand{\]}{\end{equation}}

\newcommand{\Id}{\mathcal{I}_d}
\newcommand{\ta}{\left(}
\newcommand{\tc}{\right)}
\newcommand{\para}{\parallel}

\definecolor{pine}{rgb}{0.0, 0.5, 0.0}

\allowdisplaybreaks
\begin{document}

\title{Energy-momentum tensor in QCD: \\ nucleon mass decomposition and mechanical equilibrium}
              
\author{C\'edric Lorc\'e}
\affiliation{CPHT, CNRS, Ecole Polytechnique, Institut Polytechnique de Paris, Route de Saclay, 91128 Palaiseau, France}

\author{Andreas Metz}
\affiliation{Department of Physics, SERC, Temple University, Philadelphia, PA 19122, USA}

\author{Barbara Pasquini}
\affiliation{Dipartimento di Fisica, Universit\`a degli Studi di Pavia, I-27100 Pavia, Italy}
\affiliation{Istituto Nazionale di Fisica Nucleare, Sezione di Pavia, I-27100 Pavia, Italy}

\author{Simone Rodini}
\affiliation{Dipartimento di Fisica, Universit\`a degli Studi di Pavia, I-27100 Pavia, Italy}
\affiliation{Istituto Nazionale di Fisica Nucleare, Sezione di Pavia, I-27100 Pavia, Italy}


\date{\today}

\begin{abstract}
We review and examine in detail recent developments regarding the question of the nucleon mass decomposition. We discuss in particular the virial theorem in quantum field theory and its implications for the nucleon mass decomposition and mechanical equilibrium. We  reconsider the renormalization of the QCD energy-momentum tensor in minimal-subtraction-type schemes and the physical interpretation of its components, 
as well as the role played by the trace anomaly and Poincar\'e symmetry. We also study the concept of ``quantum anomalous energy'' proposed in some works as a new contribution to the nucleon mass. Examining the various arguments, we conclude that the quantum anomalous energy is not a genuine contribution to the mass sum rule, as a consequence of translation symmetry. 
\end{abstract}

\maketitle

\section{Introduction}

In QCD the phenomenon of confinement prevents quarks and gluons to appear in the physical spectrum. Instead, one finds exclusively bound states of these elementary constituents. Hadron masses can therefore largely differ from the sum of its constituent masses. Understanding how hadron masses arise is therefore of utmost importance.

Long ago, the nucleon mass has been decomposed in a frame-independent way into a quark and a gluon contribution using the trace of the energy-momentum tensor (EMT) operator $g_{\mu\nu}T^{\mu\nu}$~\cite{Shifman:1978zn,Donoghue:1987av}. Later, a decomposition into four contributions based on the component $T^{00}$ in the rest frame has been proposed in Refs.~\cite{Ji:1994av,Ji:1995sv}. Leaving aside the precise form of the underlying renormalized operators, both decompositions are mathematically correct but provide quite different pictures of the nucleon mass, triggering debates within the hadronic physics community about their physical meaning.

In order to clarify the situation, a general Poincar\'e-covariant and scheme-independent
analysis has recently been presented in Ref.~\cite{Lorce:2017xzd} which concluded that the
above two decompositions actually mix information about mass with the constraint of mechanical
equilibrium. Keeping these two aspects of the hadronic bound state physics separated, one obtains in fact a natural decomposition of the hadron mass into a quark
contribution and a gluon contribution. Quarks being massive particles, the quark contribution
can be refined by separating the rest energy (i.e. quark mass) from the kinetic and potential
energies. A three-term decomposition of $T^{00}$ of this form is what has been discussed recently, along
with the corresponding renormalized operators in dimensional regularization (DR) in
minimal-subtraction-type (MS) schemes~\cite{Rodini:2020pis,Metz:2020vxd}. 

While agreeing with the mathematical aspects of Refs.~\cite{Lorce:2017xzd,Rodini:2020pis,Metz:2020vxd}, it is claimed in~\cite{Ji:2021mtz} that the two-term and three-term energy decompositions miss ``some fundamental insight on the origin of the nucleon mass'', namely the role played by the trace anomaly. A key step to obtain a four-term energy decomposition is to separate the EMT into traceless and trace parts, motivated by the fact that these two parts do not mix under Lorentz transformations and hence under renormalization. Focusing on $T^{00}$, it is found that the traceless part provides three quarters of the nucleon mass and the trace part provides the remaining quarter, a result referred to as a ``virial theorem'' in Refs.~\cite{Ji:1994av,Ji:1995sv,Ji:2021pys,Ji:2021mtz,Ji:2021qgo}.
Here we take a fresh look at the virial theorem in the context of quantum field theory (QFT); see also Refs.~\cite{Dudas:1991kj,Gaite:2013uqa,Davies:2020swd}.
In particular, we show that, for a closed system, the virial theorem coincides with the constraint of mechanical equilibrium put forward in Refs.~\cite{Polyakov:2002yz,Polyakov:2018zvc,Lorce:2017xzd,Lorce:2018egm}.

The present work is organized as follows. In Section~\ref{sec:virialQFT} we study in detail the QFT version of the virial theorem. After elaborating in Section~\ref{sec:interpretation} on the physical interpretation of the EMT components, we analyze in Section~\ref{sec:massanalysis} the consequences for the problem of the nucleon mass decomposition in the light of the arguments presented in Refs.~\cite{Ji:2021pys,Ji:2021mtz,Ji:2021qgo}. We review in Section~\ref{sec:renormalization} the renormalization of the EMT operators in DR in MS-type schemes and the operator structure of the different mass terms in the four-term energy decomposition, which has been under some discussion recently~\cite{Hatta:2018sqd, Tanaka:2018nae, Rodini:2020pis,Metz:2020vxd, Ji:2021pys,Ji:2021mtz,Ji:2021qgo}. We also comment on the role played by the trace anomaly and the related concept of ``quantum anomalous energy''. We show in Section~\ref{sec:lattice} the importance of preserving translation symmetry, and we argue that one should be particularly careful when providing physical interpretation to operators appearing in a lattice-regulated theory. Our findings are then summarized in Section~\ref{sec:conclusions}.
We present further details on the virial theorem and its physical meaning in various contexts in Appendix~\ref{App:virial}, and give a brief account of the DR approach in Appendix~\ref{App:DR}. 

\section{Virial theorem in Quantum Field Theory}
\label{sec:virialQFT}

The virial theorem is essentially a statement about mechanical equilibrium in a bound state, expressed as a stationarity condition on the energy under spatial dilatations. It has largely been discussed in the context of classical and quantum mechanics (see Appendix~\ref{App:virial} for a short review), but its proper transposition to field theories is less known. In this section, we present first an original derivation of the virial theorem for stationary states in QFT, and then obtain a stronger version based on the divergence of the EMT. We then discuss the relation with the plane-wave approach.

\subsection{Dilatations}\label{sec:dilatation}

In a field theory, dilatations are associated with the current\footnote{In general there can be an additional term $V^\mu$ called the virial current. It is however often possible to redefine the EMT so that the virial current does not appear.}~\cite{Callan:1970ze,Coleman:1970je}
\begin{equation}
    j^\mu_D=T^{\mu\nu}x_\nu.
\end{equation}
Note that we will not assume a priori that the system is closed, and hence that the EMT is conserved. The corresponding charge
\begin{equation}\label{Ddef}
    D=\int\ud^3x\,j^0_D=H t-G
\end{equation}
with $H=\int\ud^3x\,T^{00}$ and $G=\int\ud^3x\,T^{0i}x^i$ generates spacetime dilatations
\begin{equation}
    \phi(x)\mapsto e^{i\kappa D}\phi(x)e^{-i\kappa D}=e^{\kappa d_\phi}\phi(e^\kappa x),
\end{equation}
or, in infinitesimal form,
\begin{equation}
    \frac{1}{i}[\phi(x),D]=(x_\mu\partial^\mu+d_\phi)\phi(x),
\end{equation}
where $\phi(x)$ is a generic dynamical field appearing in the EMT and $d_\phi$ is its scale dimension. In the following, we will drop all surface terms, assuming as usual that their contributions vanish for the physical states that we consider. 

Using the Heisenberg equation of motion and the standard commutation relations with the momentum operator $P^i=\int\ud^3x\,T^{0i}$, we can write\footnote{Note that if the theory is invariant under dilatations, then $\ud D/\ud t=0$ which implies $\frac{1}{i}[H,D]=H$. If a stationary state exists, then $\langle H\rangle=0$. Eigenstates of the Hamiltonian with nonzero energy are therefore not stationary. The only possibility is that they move at the speed of light, meaning that they must be massless.}
\begin{equation}
    \frac{1}{i}[P^\mu,D]=P^\mu-g^{\mu 0}\,\frac{\ud D}{\ud t}.
\end{equation}
For $\mu=i$, this relation indicates that dilatations simply rescale the momentum. For $\mu=0$, the rescaling of the Hamiltonian is accompanied by another contribution arising from the breaking of dilatation symmetry. The latter is measured by $\ud D/\ud t$ and can be expressed as
\begin{equation}\label{dDdt}
    \begin{aligned}
    \frac{\ud D}{\ud t}&=\int\ud^3x\,\partial_0j^0_D\\
    &=\int\ud^3x\left(\partial_\mu j^\mu_D-\uvec\nabla\cdot\uvec j_D\right)\\
    &=\int\ud^3x\,T^\mu_{\phantom{\mu}\mu}+\int\ud^3x\,\mathcal F^\mu x_\mu,
    \end{aligned}
\end{equation}
where the density of four-force is defined as $\mathcal F^\mu=\partial_\lambda T^{\lambda\mu}$. Under an infinitesimal dilatation $x^\mu\mapsto(1+\delta\kappa)\,x^\mu$, the variation of the Hamiltonian is given by
\begin{equation}\label{dH}
    \begin{aligned}
    \delta H&=i[H,D]\,\delta\kappa\\
    &=\left[-\sum_i\int\ud^3x\,T^{ii}-\int\ud^3x\,\uvec{\mathcal F}\cdot\uvec x+t\,\frac{\ud H}{\ud t}\right]\delta\kappa.
    \end{aligned}
\end{equation}
Note that only spatial dilatations (i.e. those generated by $G$) matter since $[H,D]=-[H,G]$. We could therefore have directly started with $G$ instead of $D$, like in the case of point particles treated in Appendix~\ref{App:virial}, but the spirit of relativistic field theories makes it a priori more natural to consider spacetime dilatations rather than pure spatial dilatations. 

Using the operator equation~\eqref{dH}, we can write the mean variation of the energy as
\begin{equation}
     \langle\delta H\rangle=-\bar p\,\delta V-\bar f\,\delta\ell+\left\langle\frac{\ud H}{\ud t}\right\rangle\delta t,
\end{equation}
where $\langle O\rangle$ is the expectation value of the operator $O$ in some properly normalized state, $V$ is the volume and $\ell$ is the radius of a sphere containing the system,
\begin{equation}
    \bar p\equiv\frac{1}{3V}\sum_i\langle\int\ud^3x\,T^{ii}\rangle
\end{equation}
is the average isotropic stress or pressure\footnote{Since the momentum density $\langle T^{0i}\rangle(\uvec x)$ does not vanish in general, $\bar p$ includes a convective contribution and should not be confused with hydrostatic pressure which is defined in the local rest frame of the medium.}, and 
\begin{equation}
    \bar f\equiv\frac{1}{\ell}\,\langle\int\ud^3x\,\uvec{\mathcal F}\cdot\uvec x\rangle
\end{equation}
is the average radial force. The combination $\delta W=\bar p\,\delta V+\bar f\,\delta\ell$ represents the mean work exerted by the system under the infinitesimal spatial dilatation.

\subsection{Virial theorem for stationary states}\label{sec:virialthmfield}

Assuming that the Hamiltonian is time-independent, the QFT version of the virial theorem follows directly from the expectation value of Eq.~\eqref{dH} in a (normalized) \emph{stationary} state $H|\Psi\rangle=E|\Psi\rangle$
\begin{equation}\label{QFTvirial}
    \sum_i\langle\Psi|\int\ud^3x\,T^{ii}|\Psi\rangle+\langle\Psi|\int\ud^3x\,\uvec{\mathcal F}\cdot\uvec x|\Psi\rangle=0.
\end{equation}
It is a balance equation stating that in a stationary state the virtual work exerted by the system under a spatial dilatation vanishes. In other words, the system is in mechanical equilibrium. 

For a system of massive point particles, one can write 
\begin{equation}
\begin{aligned}
    \sum_iT^{ii}(x)&=\sum_k \uvec v_k(t)\cdot\uvec p_k(x)\,\delta^{(3)}(\uvec x-\uvec r_k(t)),\\
    \uvec{\mathcal F}(x)&=\sum_k\uvec F_k(t,\uvec r_a(t))\,\delta^{(3)}(\uvec x-\uvec r_k(t)),
\end{aligned}
\end{equation}
where $\uvec v_k=\ud\uvec r_k/\ud t$ is the velocity of particle $k$, $\uvec p_k$ is its momentum, and $\uvec F_k$ is the force acting on it; see Appendix~\ref{App:fieldtheory}. Noting that for a non-relativistic system the total kinetic energy is given by $\ucal T=\sum_k\frac{1}{2}\,\uvec v_k\cdot\uvec p_k$, it is easy to see that Eq.~\eqref{QFTvirial} reduces after integration to 
\begin{equation}
    \langle\Psi|\ucal T|\Psi\rangle=-\frac{1}{2}\sum_k\langle\Psi| \uvec r_k\cdot\uvec F_k|\Psi\rangle, 
\end{equation}
which is the familiar form of the virial theorem in non-relativistic quantum mechanics.

For a \emph{closed} system, the total EMT is conserved and the virial theorem reduces to
\begin{equation}\label{virialQFTclosed}
    \sum_i\langle\Psi|\int\ud^3x\,T^{ii}|\Psi\rangle=0.
\end{equation}
This relation has largely been discussed in the QED context for an electron state~\cite{Villars:1950pkp,Rohrlich:1950,Yukawa:1951,Takahashi:1952,Manoukian:1975cd,Kashiwa:1979wc}, and is a key aspect of the hadron mechanical structure~\cite{Polyakov:2002yz,Polyakov:2018zvc,Lorce:2018egm} which impacts the hadron mass decomposition~\cite{Lorce:2017xzd}. It is however usually obtained from a different approach, and therefore often not recognized as the virial theorem. The situation is different, e.g., in plasma physics~\cite{Shafranov:1966,Kulsrud:1983,Faddeev:2000qw} where Eq.~\eqref{virialQFTclosed} is well known as the virial theorem. Defining the isotropic stress or pressure distribution as $p(\uvec x)\equiv\frac{1}{3}\sum_i\langle\Psi| T^{ii}(\uvec x)|\Psi\rangle$, we see that the virial theorem for a closed system amounts simply to the von Laue condition for mechanical equilibrium~\cite{Laue:1911lrk}
\begin{equation}
    \int\ud^3 x\,p(\uvec x)=0,
\end{equation}
derived long ago in the context of classical field theory.

We observe that there exists actually some confusion in the field theory literature about the notion of virial theorem. For example, in the seminal paper~\cite{Chodos:1974je} introducing the MIT bag model two so-called ``virial theorems'' are derived using naive transpositions of the point mechanics quantity $G=\sum_k\uvec r_k\cdot\uvec p_k$ to continuum mechanics. The first one is based on $G\mapsto \Omega=\int\ud^3x\,\phi(x)\dot\phi(x)$, where $\phi(x)$ is a massless scalar field describing quarks inside the bag and $\dot\phi(x)$ is its time derivative. The second one is based on $G\mapsto\bar\Omega=\int\ud^3x\,\phi(x)\,\uvec x\cdot\uvec\nabla\dot\phi(x)$, which is close to the correct transposition $G\mapsto\int\ud^3x\,T^{0i}x^i=\int\ud^3x\,\dot\phi(x)\,\uvec x\cdot\uvec\nabla\phi(x)$. In a subsequent paper~\cite{Chodos:1974pn}, it has been observed that the key results derived from the combination of $\ud\langle\Psi|\Omega|\Psi\rangle/\ud t=0$ and $\ud\langle\Psi|\bar\Omega|\Psi\rangle/\ud t=0$ can in fact be obtained from the stationarity of the system rest energy under spatial dilatations. This variational principle expresses mechanical equilibrium and has been used later in the context of soliton models~\cite{Zahed:1986qz,Cebulla:2007ei,Polyakov:2018zvc}, where it is commonly referred to as the virial theorem. We agree with the latter naming since requiring stationarity under spatial dilatations amounts to using the correct form for the generator of spatial dilatations $G\mapsto\int\ud^3x\,T^{0i}x^i$ from which one usually derives the virial theorem $\ud\langle\Psi| G|\Psi\rangle/\ud t=0$; see Appendix~\ref{App:virial}.
\newline

Let us now consider a stronger version of the virial theorem which is most easily derived without explicit reference to dilatations. To this end, let us generalize the approach followed in Refs.~\cite{Landau:1951,Chandrasekhar:1953zz,Dudas:1991kj,Gaite:2013uqa,Polyakov:2018zvc} and write the identity
\begin{equation}\label{id1}
    \partial_0T^{0\mu}=\mathcal F^\mu-\partial_kT^{k\mu}.
\end{equation}
For a stationary state, the left-hand side vanishes using the Heisenberg equation of motion and we can write
\begin{equation}
    \partial_k\langle\Psi| T^{k\mu}|\Psi\rangle-\langle\Psi|\mathcal F^\mu|\Psi\rangle=0.
\end{equation}
Multiplying by $x^i$ and integrating over space gives for $\mu=j$
\begin{equation}
    \langle\Psi|\int\ud^3x\,T^{ij}|\Psi\rangle+\langle\Psi|\int\ud^3x\,x^i\mathcal F^j|\Psi\rangle=0.
\end{equation}
This is the QFT version of the so-called \emph{tensor} virial theorem~\cite{Parker:1954}, whose spatial trace reduces to the usual (scalar) virial theorem~\eqref{QFTvirial}. It expresses the fact that a stationary state is in mechanical equilibrium not only under isotropic dilatations, but more generally under any (infinitesimal) spatial deformation.  For a \emph{closed} system, the tensor virial theorem reduces to
\begin{equation}\label{tensorV}
    \langle\Psi|\int\ud^3x\,T^{ij}|\Psi\rangle=0.
\end{equation}

Since the virial theorem concerns only the stress tensor, one may wonder in what frame it applies. Clearly it cannot be a generic frame for it would imply that the expectation value of the total EMT must identically vanish. By a stationary state it is usually understood a normalizable state, excluding therefore momentum eigenstates. It is easy to see that the expectation value of total momentum $\uvec P$ in a stationary state vanishes, meaning that the system is in average at rest. One can indeed use e.g. the center of energy $\uvec R=\frac{1}{H}\int\ud^3x\,\uvec x\,T^{00}$ to define the position of a closed system. The velocity operator is then given by $\ud{\uvec R}/\ud t=\uvec P/H$, whose expectation value in a stationary state vanishes using again the Heisenberg equation of motion. So, in conclusion, the virial theorem simply expresses the condition of mechanical equilibrium of a massive system in the system rest frame.

\subsection{Virial theorem for momentum eigenstates}\label{sec:virialplanewave}

In particle physics, it is customary to work with four-momentum eigenstates instead of normalizable stationary states. It is actually possible to obtain in a simple way the content of the virial theorem for a closed system for such states, but the derivation turns out to involve additional information that should be distinguished from the virial theorem.

Poincar\'e symmetry implies that the forward matrix elements of the total EMT must have the form~\cite{Kobzarev:1962wt,Pagels:1966zza,Nishijima:1967byg}
\begin{equation}\label{paramEMTtot}
    \langle p|T^{\mu\nu}(x)|p\rangle=2p^\mu p^\nu,
\end{equation}
where $|p\rangle$ is covariantly normalized, i.e.~$\langle p'|p \rangle=(2\pi)^32p^0\delta^{(3)}(\uvec p'-\uvec p)$. (For simplicity, we suppress the spin labels for the nucleon states throughout this work.)
This ensures that the total four-momentum is given by
\begin{equation}\label{MomSR}
    \frac{\langle p|\int\ud^3x\,T^{0\mu}(x)|p\rangle}{\langle p|p\rangle}=p^\mu.
\end{equation}
The average total stress tensor then reads
\begin{equation}\label{Tij}
    \frac{\langle p|\int\ud^3x\,T^{ij}(x)|p\rangle}{\langle p|p\rangle}=\frac{p^ip^j}{p^0}.
\end{equation}
For a state at rest defined by $p^\mu_\text{rest}=(M,\uvec 0)$, we recover directly the tensor virial theorem for a closed system
\begin{equation}\label{virialthmprest}
    \frac{\langle p_\text{rest}|\int\ud^3x\,T^{ij}(x)|p_\text{rest}\rangle}{\langle p_\text{rest}|p_\text{rest}\rangle}=0.
\end{equation}
In the context of classical field theory, von Laue~\cite{Laue:1911lrk} showed that Eq.~(\ref{virialthmprest}) is a necessary and sufficient condition for the total four-momentum to transform as a Lorentz four-vector. The same condition must also hold in QFT~\cite{Villars:1950pkp,Yukawa:1951,Takahashi:1952} since it is just based on Lorentz symmetry.

Clearly the tensor analysis approach, which is based on Eq.~\eqref{paramEMTtot}, is very powerful and arrives at Eq.~\eqref{tensorV} in a very simple (though somewhat formal) way, with the advantage of extending its expression to any Lorentz frame. The drawback is that it keeps the physical meaning obscure, and in particular the fact that it includes automatically the virial theorem which is associated with spatial dilatations. To clarify the physical meaning of Eq.~\eqref{paramEMTtot}, we first note that the tensor virial theorem for a closed system in Eq.~\eqref{virialthmprest} can be expressed in an arbitrary frame as
\begin{equation}\label{covQFTvirialexpr}
    (u^\mu u_\alpha-\delta^\mu_\alpha)(u^\nu u_\beta-\delta^\nu_\beta)\,\frac{\langle p|\int\ud V\,T^{\alpha\beta}(x)|p\rangle}{\langle p|p\rangle}=0,
\end{equation} 
where $u^\mu=p^\mu/M$ is the system four-velocity and $\ud V=u^0\ud^3x$ is the Lorentz-invariant proper volume element. It implies that
\begin{equation}
    \frac{\langle p|\int\ud V\,T^{\mu\nu}(x)|p\rangle}{\langle p|p\rangle}=\frac{\langle p|T^{\mu\nu}(0)|p\rangle}{2M}=A\, u^\mu u^\nu,
\end{equation}
where we used translation invariance and the fact that $\langle p|p\rangle=(2\pi)^32p^0\delta^{(3)}(\uvec 0)=2M\,(2\pi)^3u^0\delta^{(3)}(\uvec 0)=2M\int\ud V$. Note that one arrives at the same conclusion using tensor analysis and the conservation of the total EMT which excludes other possible Lorentz structures involving $g^{\mu\nu}$ or polarization tensors~\cite{Cotogno:2019vjb}. We stress that the virial theorem does not require to know the coefficient $A$. The latter is fixed by the additional requirement that the proper energy is the mass of the system
\begin{equation}\label{massconstraint}
    u_\mu u_\nu\,\frac{\langle p|\int\ud V\,T^{\mu\nu}(x)|p\rangle}{\langle p|p\rangle}=M
\end{equation}
or, equivalently, by the requirement of four-momentum conservation~\eqref{MomSR}. This implies that $A=M$, leading us back to Eq.~\eqref{paramEMTtot}. This analysis shows clearly that the expression~\eqref{paramEMTtot} combines in fact \emph{two} distinct physical aspects of bound systems: one is the virial theorem expressing mechanical equilibrium~\eqref{covQFTvirialexpr} and the other is that the mass of the system is $M$~\eqref{massconstraint}.

\section{EMT matrix elements and their interpretation}\label{sec:interpretation}

Now that the QFT version of the virial theorem is well identified, we would like to add further discussion about the physical interpretation of the EMT matrix elements. For our purpose, it will be sufficient to work with the symmetric (or Belinfante) form of the EMT. We will also assume that the total EMT can be written as the sum of partial EMTs
\begin{equation}\label{partialEMT}
    T^{\mu\nu}(x)=\sum_a T^{\mu\nu}_a(x)
\end{equation}
associated with the individual species of constituents in the system. In QCD, we will typically separate the system into quark and gluon contributions. The quark contribution can further be decomposed into flavor contributions. Note that vacuum expectation values are always implicitly subtracted from these operators.

\subsection{Parametrization}\label{sec:paramEMT}

For a spin-$1/2$ target, the matrix elements of the symmetric (or Belinfante) EMT can be parametrized in general as~\cite{Kobzarev:1962wt, Pagels:1966zza, Ji:1996ek} 
\begin{equation}
    \langle p'|T^{\mu\nu}_a(0)|p \rangle=\overline u(p')\Gamma^{\mu\nu}_a(P,\Delta)u(p),
\end{equation}
with
\begin{equation}
    \Gamma^{\mu\nu}_a(P,\Delta)=\frac{P^{\{\mu}\gamma^{\nu\}}}{M}\,A_a(\Delta^2)+\frac{P^{\{\mu}i\sigma^{\nu\}\lambda}\Delta_\lambda}{2M}\,B_a(\Delta^2)+\frac{\Delta^\mu\Delta^\nu-g^{\mu\nu}\Delta^2}{M}\,C_a(\Delta^2)+Mg^{\mu\nu}\bar C_a(\Delta^2),
\end{equation}
where $a^{\{\mu}b^{\nu\}}=\frac{1}{2}(a^\mu b^\nu+a^\nu b^\mu)$, $P=\frac{1}{2}(p'+p)$ is the average four-momentum, $\Delta=p'-p$ is the four-momentum transfer, and $a$ is just a generic label specifying the EMT contribution. 
The gravitational form factors depend on $\Delta^2$ only and therefore are frame-independent.
In the forward limit, this parametrization reduces to
\begin{equation}\label{forwardAmpl}
    \langle p |T^{\mu\nu}_a(0)|p \rangle = 2p^\mu p^\nu A_a(0)+2M^2g^{\mu\nu}\bar C_a(0) .
\end{equation}

\subsection{Spatial distributions}

The expectation value of the EMT tensor in some physical nucleon state $|\Psi\rangle$ (not necessarily stationary) at time $t=0$ is given by
\begin{equation}\label{Tdensity}
    \langle\Psi|T^{\mu\nu}_a(0,\uvec x)|\Psi\rangle=\int\frac{\ud^3p'}{(2\pi)^3}\,\frac{\ud^3p}{(2\pi)^3}\,\tilde\Psi^*(\uvec p')\tilde\Psi(\uvec p)\,e^{-i\uvec(\uvec p'-\uvec p)\cdot\uvec x}\,\frac{\langle p'|T^{\mu\nu}_a(0)|p\rangle}{\sqrt{2p'^02p^0}},
\end{equation}
where the momentum-space wave packet is defined as $\tilde\Psi(\uvec p)=\langle p|\Psi\rangle/\sqrt{2p^0}$. Applying a Wigner transform, this can be rewritten in a phase-space representation as~\cite{Lorce:2018zpf,Lorce:2018egm,Lorce:2021gxs}
\begin{equation}
    \langle\Psi|T^{\mu\nu}_a(0,\uvec x)|\Psi\rangle=\int\frac{\ud^3P}{(2\pi)^3}\,\ud^3R\,\rho_\Psi(\uvec R,\uvec P)\,\langle T^{\mu\nu}_a\rangle(\uvec x-\uvec R,\uvec P)
\end{equation}
with
\begin{equation}
\rho_\Psi(\uvec R,\uvec P)\equiv\int\frac{\ud^3q}{(2\pi)^3}\,e^{-i\uvec q\cdot\uvec R}\,\tilde\Psi^*(\uvec P+\tfrac{\uvec q}{2})\tilde\Psi(\uvec P-\tfrac{\uvec q}{2})
\end{equation}
the nucleon Wigner distribution and
\begin{equation}\label{intEMT}
    \langle T^{\mu\nu}_a\rangle(\uvec r,\uvec P)\equiv\int\frac{\ud^3\Delta}{(2\pi)^3}\,e^{-i\uvec\Delta\cdot\uvec r}\,\frac{\langle P+\tfrac{\Delta}{2}|T^{\mu\nu}_a(0)|P-\tfrac{\Delta}{2}\rangle}{\sqrt{2p'^02p^0}}
\end{equation}
the internal EMT distribution for a nucleon localized in the Wigner sense in phase space. The average rest frame $\uvec P=\uvec 0$ is known as the Breit frame, where Eq.~\eqref{intEMT} reduces to the 3D EMT distributions introduced in Ref.~\cite{Polyakov:2002yz} and reviewed in Ref.~\cite{Polyakov:2018zvc}. When $\uvec P\neq\uvec 0$ one can choose without loss of generality the $z$ axis along $\uvec P$. Integrating over $r_z$, one obtains the 2D EMT distributions which become genuine probabilistic distributions in the infinite-momentum limit~\cite{Lorce:2017wkb,Lorce:2018egm,Freese:2021czn}. In these two cases there is no energy transfer, $p'^0=p^0$, so that the spatial distributions are in fact static, i.e., time-independent. Relativistic spatial charge distributions are constructed in a similar way using the charge current $j^\mu(x)$~\cite{Lorce:2020onh}.

\subsection{Four-momentum sum rules}

Integrating Eq.~\eqref{Tdensity} over all space and using the parametrization~\eqref{forwardAmpl}, we arrive at
\begin{equation}
     \langle\Psi|\int\ud^3x\,T^{\mu\nu}_a(0,\uvec x)|\Psi\rangle=\int\frac{\ud^3p}{(2\pi)^3}\,|\tilde\Psi(\uvec p)|^2\,\frac{p^\mu p^\nu A_a(0)+M^2g^{\mu\nu}\bar C_a(0)}{p^0}.
\end{equation}
For a state with well-defined momentum $\uvec p$, this reduces to
\begin{equation}\label{forwardT}
     \frac{\langle p|\int\ud^3x\,T^{\mu\nu}_a(0,\uvec x)|p \rangle}{\langle  p |p \rangle}=\frac{p^\mu p^\nu A_a(0)+M^2g^{\mu\nu}\bar C_a(0)}{p^0},
\end{equation}
which is also valid for $t\neq 0$ because $|p \rangle$ is an energy eigenstate. With the four-momentum operator being defined as
\begin{equation}
    P^\mu_a(t)=\int\ud^3x\,T^{0\mu}_a(t,\uvec x),
\end{equation}
we recover from Eq.~\eqref{forwardT} the expression~\cite{Ji:1997pf}
\begin{equation}\label{forwardPop}
     \frac{\langle p |P^\mu_a(t)|p \rangle}{\langle  p |p \rangle}=p^\mu A_a(0)+\frac{M^2}{p^0}\,g^{0\mu}\bar C_a(0).
\end{equation}
Summing over all the contributions we should recover the four-momentum of the state, leading therefore to a momentum sum rule for $\mu=i$
\begin{equation}
    \sum_a A_a(0)=1
\end{equation}
and an energy sum rule for $\mu=0$
\begin{equation}\label{SRenergy}
    \sum_a \left[A_a(0)+\frac{M^2}{(p^0)^2}\,\bar C_a(0)\right]=1.
\end{equation}
Equation~(\ref{SRenergy}) must be true in any frame and we therefore conclude that
\begin{equation}\label{SRbarC}
    \sum_a\bar C_a(0)=0.
\end{equation}

\subsection{Lorentz symmetry and physical interpretation}

From the point of view of Lorentz symmetry, it is interesting to decompose the EMT into a symmetric traceless contribution and a trace contribution~\cite{Ji:1994av,Ji:1995sv}, since these belong to different representations of the Lorentz group and hence do not mix under Lorentz transformations. One can then rewrite Eq.~\eqref{forwardT} as
\begin{equation}\label{repdec}
     \frac{\langle p|\int\ud^3x\,T^{\mu\nu}_a(x)|p \rangle}{\langle  p|p \rangle}=\frac{1}{p^0}\left[\left(p^\mu p^\nu-\frac{M^2}{4}\,g^{\mu\nu}\right) A_a(0)+\frac{M^2}{4}\,g^{\mu\nu}\left(A_a(0)+4\bar C_a(0)\right)\right].
\end{equation}
Alternatively, we may note that the four-momentum of the system is a timelike four-vector that can be used to provide a natural foliation of spacetime into spacelike hypersurfaces. From the physical point of view, this means that $p^\mu$ specifies in a covariant way the rest frame of the system. We can then decompose the EMT into parallel and orthogonal contributions to $p^\mu$~\cite{Eckart:1940te,Bergmann:1976,Smarr:1977uf,Lorce:2017xzd}
\begin{equation}\label{pfoldec}
     \frac{\langle p |\int\ud^3x\,T^{\mu\nu}_a(x)|p \rangle}{\langle  p|p\rangle}=\frac{1}{p^0}\left[p^\mu p^\nu\left(A_a(0)+\bar C_a(0)\right)-\left(p^\mu p^\nu-M^2g^{\mu\nu}\right)\bar C_a(0)\right].
\end{equation}
In the rest frame, we have in particular
\begin{equation}\label{restpic}
    \frac{\langle p_\text{rest}|\int\ud^3x\,T^{\mu\nu}_a(x)|p_\text{rest}\rangle}{\langle  p_\text{rest}|p_\text{rest}\rangle}=M\begin{pmatrix}A_a(0)+\bar C_a(0)&0&0&0\\
    0&-\bar C_a(0)&0&0\\
    0&0&-\bar C_a(0)&0\\
    0&0&0&-\bar C_a(0)\end{pmatrix},
\end{equation}
which shows that the combination $A_a(0)+\bar C_a(0)$ represents the fraction of the system rest energy carried by the subsystem $a$, and that $-\bar C_a(0) M$ represents the rest isotropic stress of this subsystem integrated over the volume. Dividing by the proper volume $V$ of the system, one can interpret $A_a(0)+\bar C_a(0)$ as the average energy density and $-\bar C_a(0)$ as the average isotropic pressure, both defined in the rest frame of the system and expressed in units of $M/V$~\cite{Lorce:2017xzd}. 

Thanks to Eq.~\eqref{restpic}, we can easily understand the physical meaning of the sum rule~\eqref{SRbarC}. It is simply the virial theorem derived in Section~\ref{sec:virialQFT}, expressing the mechanical equilibrium of the system~\cite{Lorce:2017xzd}. The fact that we have in the forward limit two gravitational form factors $A_a(0)$ and $\bar C_a(0)$ satisfying two \emph{independent} sum rules~\eqref{SRenergy} and~\eqref{SRbarC} strongly suggests that they correspond to two \emph{distinct} aspects of the physics of bound states. This is further motivated by the structure of the EMT which adopts its simplest form in the nucleon rest frame~\eqref{restpic}.
\newline

Although the habit of interpreting expectation values of the EMT using the language of continuum mechanics has a long history~\cite{Pais:1949vdk,Villars:1950pkp,Rohrlich:1950,Yukawa:1951,Takahashi:1952,Kashiwa:1979wc,Shafranov:1966,Kulsrud:1983,Faddeev:2000qw,Polyakov:2002yz,Milton:2016sev,Burkert:2018bqq,Shayit:2021kgn}, some concerns about this picture have recently been expressed in Ref.~\cite{Ji:2021mtz}. It is claimed that the interpretation in terms of energy density and pressure makes sense only under some conditions. One of them is that the particles mean free paths must be much smaller than the volume elements. It is then concluded that ``in QCD, only at high-temperature and density, a fluid description of the combined quark and gluon plasma might make sense''~\cite{Ji:2021mtz}. We remind however that the alluded conditions indicate in fact when a \emph{macroscopic} continuum description can be used when the \emph{microscopic} degrees of freedom are discrete. They tell us, e.g., under what circumstances one can describe a gas composed of a large number of particles as a classical continuous medium from an effective macroscopic point of view. 

In QFT the fundamental degrees of freedom are \emph{fields}, and particles emerge as somewhat localized excitations of the latter. Unlike classical field theory, which is usually seen as an \emph{effective} macroscopic description, QFT is a \emph{fundamental} continuous description. Following quantum mechanics, the expectation value $\langle \Psi|T^{\mu\nu}(x)|\Psi\rangle$ simply represents the quantum mean value of the EMT at some spacetime point $x$. There is no coarse graining\footnote{In reality, contributions to the EMT usually diverge in perturbation theory in terms of the bare degrees of freedom, so that the theory has to be renormalized. In some sense renormalization is akin to coarse graining since it amounts to defining effective finite degrees of freedom through the choice of particular renormalization conditions and introducing a so-called renormalization scale.} involved and one can safely apply the language of continuum mechanics. We also point out that the average energy density and pressure are here defined by dividing the total rest energy and work by the \emph{whole} proper volume of the nucleon, which is necessarily larger than the quark and gluon mean free paths. The conditions of applicability of the effective macroscopic description are therefore also satisfied.

Let us illustrate this with an example. If we consider an ideal gas from a macroscopic point of view, the stress tensor will contain both a convective contribution and an internal pressure contribution, but from the microscopic point-particle perspective both arise from the motion of the particles and hence are purely convective. The macroscopic distinction arises simply because of the coarse graining procedure, which defines an effective local pressure by averaging over distances larger than the particle mean free path. In QFT, the situation is reversed since the microscopic degrees of freedom are not particles but quantum fields. So on top of the convective contribution (i.e., kinetic energy), the microscopic stress tensor will in general also receive some internal contribution. For a stationary state there cannot be friction, so the internal stress is akin to a conservative potential energy and can accordingly be interpreted as pressure. More precisely, average pressure\footnote{In this work we are talking about the non-thermal part of energy density and pressure. The thermal part is defined in finite temperature field theory from the thermal average of the same components $T^{00}$ and $T^{ii}$ of the EMT~\cite{Meyer:2007fc,Cheng:2007jq,Bazavov:2009zn}. The non-thermal part has recently been investigated in lattice field theory in Refs.~\cite{Shanahan:2018nnv,Yanagihara:2018qqg,Yanagihara:2019foh,Yanagihara:2020tvs}.} is simply understood in the sense of $\bar p=-\delta E/\delta V$ where $\delta E=\langle\Psi|\delta H|\Psi\rangle$ is the variation of energy associated with an infinitesimal change of volume $\delta V$, see Section~\ref{sec:dilatation}.

\section{Tensor analysis of nucleon mass}\label{sec:massanalysis}

Having discussed in detail the physical interpretation of EMT matrix elements, we now address specifically the question of the nucleon mass decomposition. We will adopt here the  approach of Ref.~\cite{Lorce:2017xzd} which extends the work of Polyakov~\cite{Polyakov:2002yz} to the non-conserved (partial) EMTs of the partons. It is very general in the sense that it is based only on the components of the EMT, and not on the particular form assumed by the latter in a given theory and renormalization scheme. Note that this does not mean of course that the magnitude of the individual contributions are renormalization-scheme independent. Again, we will consider that the total EMT can be written as a sum of partial EMTs as in Eq.~\eqref{partialEMT}.

\subsection{Proper energy decomposition}\label{sec:properenergy}

In special relativity, mass is defined by the equation $p^\mu p_\mu=M^2$, where $p^\mu$ is the total four-momentum of the system. In QFT, this becomes an operator identity 
\begin{equation}
    P^\mu P_\mu=M^2,
\end{equation}
with $P^\mu$ the total four-momentum operator. Since we can write 
\begin{equation}
M|p\rangle=\frac{P^\mu P_\mu}{M}\,| p\rangle=P^\mu| p\rangle\,u_\mu
\end{equation}
for a massive momentum eigenstate, we conclude that (invariant) mass is fundamentally the proper energy of the system~\cite{J.L.Synge:1960zz,Lorce:2017xzd}, i.e. the Lorentz-invariant expression of the rest-frame energy\footnote{Similarly, proper time and proper length are the Lorentz-invariant expressions of rest-frame lapse and distance.}. 

A mass decomposition is therefore a proper energy decomposition. Using Eq.~\eqref{forwardPop}, it is given by the Lorentz-invariant relation
\begin{equation}\label{massdeccov}
     M=\sum_a\frac{\langle p|\int\ud^3x\,T^{0\mu}_a(x)|p\rangle}{\langle p|p\rangle} \,u_\mu=\sum_a\left[A_a(0)+\bar C_a(0)\right]M.
\end{equation}
In the rest frame, the nucleon four-velocity reduces to $u^\mu=(1,\uvec 0)$ and the proper energy $P^\mu u_\mu$ coincides with the energy $P^0=\int\ud^3x\,T^{00}(x)$, which was the starting point of Refs.~\cite{Ji:1994av,Ji:1995sv}. Nucleons being composed of quarks and gluons, it has been argued in Ref.~\cite{Lorce:2017xzd} that a natural decomposition of the nucleon mass will consist of two terms,
\begin{equation}\label{CLdec}
    M=U_q+U_g,
\end{equation}
where 
\begin{equation}
    U_a\equiv\left[A_a(0)+\bar C_a(0)\right]M
\end{equation}
is interpreted as the internal proper energy associated with parton species $a$. 

It is possible to refine this decomposition.
An obvious and theoretically trivial refinement would be to separate $U_q$ into contributions from the different quark flavors. We do not elaborate on this point which only matters when studying numerical values for the contributions to the nucleon mass. Let us rather focus on another refinement.
Since quarks are massive particles we can write~\cite{Lorce:2017xzd, Rodini:2020pis, Metz:2020vxd} 
\begin{equation}\label{MRPdec}
    M=(U_q-U_m)+U_m+U_g,
\end{equation}
where
\begin{equation}
 U_m\equiv\frac{\langle p|\int\ud V\,(\overline\psi m\psi)(x)| p\rangle}{\langle p|p\rangle}=  \frac{\langle p|\int\ud^3x\,(\overline\psi m\psi)(x)| p\rangle}{\langle p|p\rangle}\, u^0 
\end{equation} can be interpreted as the quark mass or rest-energy contribution, and therefore $(U_q-U_m)$ as the quark proper kinetic and potential energies. This is motivated by the familiar decomposition of a free-particle energy into kinetic and rest energy contributions,
\begin{equation}
    \sqrt{\uvec p^2+M^2}=(\sqrt{\uvec p^2+M^2}-M)+M.
\end{equation}
The three-term energy decomposition~\eqref{MRPdec} has recently been obtained in DR~\cite{Rodini:2020pis,Metz:2020vxd}, and will be discussed in more detail in Section~\ref{sec:renormalization}. 

The pioneering four-term energy decomposition proposed in Refs.~\cite{Ji:1994av,Ji:1995sv} has recently been slightly reorganized in Refs.~\cite{Bali:2016lvx,Yang:2018nqn,Liu:2021gco,Ji:2021mtz}. To obtain its modern form within the present tensor analysis approach, we need to write
\begin{equation}\label{Jirefinement}
M=M_q+M_g+M_m+M_\text{a},
\end{equation}
using the refinement
\begin{equation}\label{MqMgdef}
    \begin{aligned}
       M_q&\equiv U_q-c_m M_m-c_\text{a} M_\text{a},\\
       M_g&\equiv U_g-(1-c_m) M_m-(1-c_\text{a}) M_\text{a},\\
       M_m&\equiv U_m,\\
       M_\text{a}&\equiv\frac{1}{4}\,\frac{\langle p|\int\ud V\,\big[\frac{\beta}{2g}\,F^2+\gamma_m\overline\psi m\psi\big](x)| p\rangle}{\langle p|p\rangle},
    \end{aligned}
\end{equation}
with $c_{m,\text{a}}$ two renormalization-scheme-dependent coefficients, $\gamma_m$ the quark mass anomalous dimension, $\beta$ the QCD beta function, and $F^{\mu\nu}$ the gluon field strength tensor. From the tensor analysis perspective, we are unable to find any motivation for interpreting $M_{q,g}$ as the quark/gluon kinetic and potential energies, as was proposed in Refs.~\cite{Ji:1994av,Ji:1995sv}. Without a clear physical interpretation of $M_q$ and $M_g$, the refinement~\eqref{Jirefinement} appears somewhat ad hoc, leading to the conclusion in Ref.~\cite{Lorce:2017xzd} that the introduction of $M_m$ and $M_\text{a}$ in the nucleon mass decomposition is in some sense arbitrary.

Contrary to the analysis of Ref.~\cite{Lorce:2017xzd}, the derivation of the four-term energy decomposition of Refs.~\cite{Ji:1994av,Ji:1995sv} did actually not start from a decomposition of the total EMT into quark and gluon contributions. Instead, the total EMT is first decomposed into
\begin{equation}\label{JiTdec}
    T^{\mu\nu}=\bar T^{\mu\nu}+\hat T^{\mu\nu},
\end{equation}
where the traceless and trace parts are defined as
\begin{equation}\label{JiTdecdef}
    \begin{aligned}
       \bar T^{\mu\nu}&\equiv T^{\mu\nu}-\frac{1}{4}\,g^{\mu\nu}T^\lambda_{\phantom{\lambda}\lambda},\\
       \hat T^{\mu\nu}&\equiv\frac{1}{4}\,g^{\mu\nu}T^\lambda_{\phantom{\lambda}\lambda}.
    \end{aligned}
\end{equation}
Working for convenience in the rest frame, the proper energy density is simply given by the $\mu=\nu=0$ component so that
\begin{equation}\label{compmix}
    T^{00}=\left(\frac{3}{4}\,T^{00}+\frac{1}{4}\sum_i T^{ii}\right)+\left(\frac{1}{4}\,T^{00}-\frac{1}{4}\sum_i T^{ii}\right)
\end{equation}
following the decomposition in Eq.~\eqref{JiTdec}. Using now the virial theorem~\eqref{virialthmprest} and the definition of mass~\eqref{massdeccov}, one concludes that the so-called ``tensor'' and ``scalar'' energies are given by
\begin{equation}\label{virialdec}
\begin{aligned}
     E_T\equiv\frac{\langle p_\text{rest}|\int\ud^3x\,\bar T^{00}(x)|p_\text{rest}\rangle}{\langle p_\text{rest}|p_\text{rest}\rangle}&=\frac{3}{4}\,M,\\
    E_S\equiv\frac{\langle p_\text{rest}|\int\ud^3x\,\hat T^{00}(x)|p_\text{rest}\rangle}{\langle p_\text{rest}|p_\text{rest}\rangle}&=\frac{1}{4}\,M.
    \end{aligned}
\end{equation}

In Refs.~\cite{Ji:1994av,Ji:1995sv} the result~\eqref{virialdec} was obtained using Eq.~\eqref{paramEMTtot} and was interpreted as \emph{analogous} to the virial theorem. In recent papers~\cite{Ji:2021pys,Ji:2021mtz,Ji:2021qgo}, the relation 
\begin{equation}\label{Jivirial}
    E_T=3\,
    E_S
\end{equation}
is now referred to as the ``relativistic virial theorem'', motivated by the observation that one can deduce from it the familiar non-relativistic expression $2\langle\Psi|\ucal T|\Psi\rangle=-\langle\Psi|\ucal V|\Psi\rangle$ in the case of the positronium system in Coulomb gauge. Strictly speaking, this argument does not prove that Eq.~\eqref{Jivirial} is the actual relativistic virial theorem. It only indicates that Eq.~\eqref{Jivirial} holds true \emph{because of} the relativistic virial theorem. As stressed earlier, there is some confusion in the literature about what the virial theorem is in field theory. In order to clarify this point, we reviewed in Appendix~\ref{App:virial} its derivation in both classical and quantum mechanics, and we extended explicitly the derivation to QFT in Section~\ref{sec:virialQFT}. The result for a closed system is given in Eq.~\eqref{virialthmprest}, and expresses the fact that the system is in mechanical equilibrium~\cite{Polyakov:2002yz,Polyakov:2018zvc,Lorce:2017xzd,Lorce:2018egm}. As shown in Section~\ref{sec:virialplanewave}, the virial theorem is already contained in Eq.~\eqref{paramEMTtot} used to derive the relation~\eqref{Jivirial}. The latter should therefore be considered as a \emph{corollary} of the relativistic virial theorem~\eqref{virialthmprest} rather than the virial theorem per se. Indeed, using the decomposition~\eqref{JiTdec} and the virial theorem we have
\begin{equation}
    \sum_i\frac{\langle p_\text{rest}|\int\ud^3x\,\bar T^{ii}(x)|p_\text{rest}\rangle}{\langle p_\text{rest}|p_\text{rest}\rangle}=-\sum_i\frac{\langle p_\text{rest}|\int\ud^3x\,\hat T^{ii}(x)|p_\text{rest}\rangle}{\langle p_\text{rest}|p_\text{rest}\rangle}.
\end{equation}
Now, from the definition of the traceless and trace parts~\eqref{JiTdecdef} it follows that
\begin{equation}
    \begin{aligned}
       \sum_i\bar T^{ii}(x)&=\bar T^{00}(x),\\
       \sum_i\hat T^{ii}(x)&=-3\,\hat T^{00}(x),
    \end{aligned}
\end{equation}
leading then to Eq.~\eqref{Jivirial}.

While the decomposition~\eqref{JiTdec} can certainly be motivated by the fact that $\bar T^{\mu\nu}$ and $\hat T^{\mu\nu}$ do not mix under renormalization and Lorentz transformations, it comes at the price of mixing $T^{00}$ and $T^{ii}$ components, as clearly indicated by Eq.~\eqref{compmix}. We have seen that the Lorentz-invariant definition of mass~\eqref{massdeccov} requires only the four-momentum density operator $T^{0\mu}$. The stress tensor $T^{ij}$ has nothing to do with mass, so introducing it in the mass decomposition means that one is mixing information about the proper energy content with the requirement of mechanical equilibrium~\cite{Lorce:2017xzd}. This can be seen directly from the parametrization of the modern\footnote{In the original works~\cite{Ji:1994av,Ji:1995sv}, the contribution arising from the anomalous quark mass dimension is attributed to $H_m=\int\ud^3x\,\hat T^{00}_m(x)$, while in the modern form it is part of $H_\text{a}=\int\ud^3x\,\hat T^{00}_\text{a}(x)$.} four-term energy decomposition
\begin{equation}
    \begin{aligned}
       M_q&=\frac{3}{4}\left(a-\frac{b}{1+\gamma_m}\right)M,\\
       M_g&=\frac{3}{4}\left(1-a\right)M,\\
       M_m&=\frac{b}{1+\gamma_m}\,M,\\
       M_\text{a}&=\frac{1}{4}\left(1-\frac{b}{1+\gamma_m}\right)M.
    \end{aligned}
\end{equation}
There are only two unknown numbers $a$ and $b$ (known as Ji's parameters) for four terms. There must therefore be two independent linear relations. One is obviously the mass sum rule~\eqref{Jirefinement}. The other independent relation is 
\begin{equation}
    M_q+M_g=3 M_\text{a},
\end{equation}
which follows from the virial theorem~\eqref{virialthmprest} and the definition of the $a$ and $b$ parameters\footnote{Ji's parameters can be expressed in terms of the EMT form factors introduced in Section~\ref{sec:paramEMT}. We have simply $a=A_q(0)$, while the relation for $b$ is more complicated and depends on the renormalization scheme, see the discussion in Section~\ref{sect:renormalization}.} given in Refs.~\cite{Ji:1994av,Ji:1995sv}
\begin{equation}
    \begin{aligned}
       \langle p|\bar T^{\mu\nu}_q(0)|p\rangle&=2a\left(p^\mu p^\nu-\frac{1}{4}\,g^{\mu\nu}M^2\right),\\
       (1+\gamma_m)\langle p|(\overline\psi m\psi)(0)| p\rangle&=2b\,M^2.
    \end{aligned}
\end{equation}

To sum up, there are \emph{two} independent pieces of information encoded in the forward matrix elements of the EMT. One is the mass of the system encoded in the rest frame by the $T^{00}$ component. The other is the virial theorem, which expresses the mechanical equilibrium of the system, encoded in the rest frame by the $T^{ij}$ components. A genuine mass decomposition should not mix these two aspects. This is the case of the two-term energy decomposition~\eqref{CLdec} proposed in Ref.~\cite{Lorce:2017xzd} and its three-term refinement~\eqref{MRPdec} found in~\cite{Rodini:2020pis,Metz:2020vxd}.

\subsection{The role of the trace anomaly in the origin of the nucleon mass}\label{sec:reltracemass}

In a response to criticisms of the four-term energy decomposition, in Ref.~\cite{Ji:2021mtz} it was argued that ``it is unclear what new insight can be brought to the understanding of the proton mass through the process of regrouping if
any. To the contrary, this rearrangement stands to lose much''. It was concluded that the two-term and three-term energy decompositions miss ``the fundamental insight on the origin of the proton mass''. By ``fundamental insight'', we believe the role played by the trace anomaly in the nucleon mass was meant. In this subsection we address this important point.

The trace of the EMT measures the breaking of dilatation symmetry due to the presence of mass scales in the theory. Mass scales are generally provided at the classical level by the constituent masses. At the quantum level, another scale appears through the process of renormalization leading to anomalous contributions to the trace of the EMT. 

At the operator level, there is no connection between the trace of the EMT $T^\mu_{\phantom{\mu}\mu}$ and the mass of a physical state $M^2=P^\mu P_\mu$ with $P^\mu=\int\ud^3x\,T^{0\mu}(x)$. The connection can however be made at the level of expectation values

in a stationary state $|\Psi\rangle$ thanks to the virial theorem $\langle\Psi|\int\ud^3x\,T^{ij}(x)|\Psi\rangle=0$ which implies~\cite{Manoukian:1975cd,Kashiwa:1979wc}
\begin{equation}\label{traceSR}
 M=\langle\Psi|\int\ud^3x\,T^\mu_{\phantom{\mu}\mu}(x)|\Psi\rangle.   
\end{equation}
In the literature the relation is in fact usually derived in terms of four-momentum eigenstates directly from the trace of Eq.~\eqref{paramEMTtot}, leading to~\cite{Shifman:1978zn,Donoghue:1987av}
\begin{equation}
    \langle p|T^\mu_{\phantom{\mu}\mu}(0)|p\rangle=2M^2.
\end{equation}
This derivation is manifestly Lorentz invariant but hides the fact that it implicitly makes use of the virial theorem, since Eq.~\eqref{paramEMTtot} can be obtained solely from Poincar\'e symmetry arguments. As a result of this relation, there seems to be a deeply-rooted idea that the trace anomaly must fundamentally be connected to the nucleon mass, and so one might expect it to appear \emph{explicitly} in the mass budget. We stress however that this appearance can only be made through the use of the virial theorem, since matrix elements of $\sum_iT^{ii}$ are necessarily involved. The virial theorem is useful because it allows one to relate the average value of different quantities, like e.g. the average kinetic and potential energies in point mechanics, and therefore to reexpress the total energy in a different way, see Appendix~\ref{App:virialMQ} for an example in Dirac theory. However, it does not provide any clue about the \emph{actual} origin of mass.

One can also understand this from the point of view of dilatations. Note that the EMT can generally be interpreted as the response of the system to infinitesimal spacetime distortions $x^\mu\mapsto x^\mu+\varepsilon\xi^\mu(x)$~\cite{DiFrancesco:1997nk}. Indeed, the corresponding variation of the action is given by
\begin{equation}
\delta S=\varepsilon\int\ud^4x\,T^{\mu\nu}\partial_\mu\xi_\nu.    
\end{equation}
When the EMT is symmetric, we can write 
\begin{equation}
    \delta S=-\frac{1}{2}\int\ud^4x\,T^{\mu\nu}\delta g_{\mu\nu}
\end{equation}
because an infinitesimal spacetime distortion can be seen as a diffeomorphism under which the variation of the (symmetric) metric is given by $\delta g_{\mu\nu}=-\varepsilon (\partial_\mu\xi_\nu+\partial_\nu\xi_\mu)$. It follows that temporal dilatations tell us something about the Hamiltonian and hence the mass of the system, while spatial dilatations lead to the virial theorem and tell us something about mechanical equilibrium. The trace anomaly, which is associated with isotropic spacetime dilatations, then necessarily combines these two independent aspects of a bound system.

So, we do not consider that writing
\begin{equation}
    M=\langle\Psi|\int\ud^3x\,T^{00}(x)|\Psi\rangle=\langle\Psi|\int\ud^3x\left(T^{00}-\frac{1}{4}\,T^\mu_{\phantom{\mu}\mu}\right)\!(x)|\Psi\rangle+\frac{1}{4}\,\langle\Psi|\int\ud^3x\,T^\mu_{\phantom{\mu}\mu}(x)|\Psi\rangle\label{eq:mass-definition}
\end{equation}
brings any fundamental insight into the question of the origin of the nucleon mass~\cite{Lorce:2017xzd}. It is just a particular case of the more general relation
\begin{equation}\label{alphadec}
    M=\langle\Psi|\int\ud^3x\left(T^{00}- \alpha\,T^\mu_{\phantom{\mu}\mu}\right)\!(x)|\Psi\rangle+\alpha\,\langle\Psi|\int\ud^3x\,T^\mu_{\phantom{\mu}\mu}(x)|\Psi\rangle
\end{equation}
which is obviously true for any value of $\alpha$, independently of the virial theorem.

We repeat that for $\alpha\neq 0$ the two terms in Eq.~\eqref{alphadec} correspond to different combinations of EMT components, which is not a natural thing to do from the tensor analysis perspective since the individual terms would then correspond to different physical quantities rather than different contributions to the same physical quantity~\cite{Lorce:2017xzd}. This is similar to what happens, e.g., in thermodynamics. One can define the notion of enthalpy $H$ as the sum of internal energy $U$ and pressure-volume work $W=pV$. However, one does usually not consider that the relation $U=H-pV = (U + pV) - pV$ represents an actual decomposition of internal energy.

As a final remark, we find it somewhat surprising that in Section IV of Ref.~\cite{Ji:2021mtz} it is argued that using the relation~\eqref{Tij} for the first term of Eq.~\eqref{alphadec} in the case $\alpha=1$ does not make a lot of sense physically, while in Section II of the same paper the relation~\eqref{Tij} is used in the case $\alpha=1/4$ to provide some alleged fundamental insight, namely the fact that $1/4$ of the nucleon mass comes from the EMT trace. Once again, the main motivation for the choice $\alpha=1/4$ is that the two terms do not mix with each other under Lorentz transformations and renormalization, but \emph{without} the virial theorem these terms have separately no clear relation to mass. Only the two-term~\eqref{CLdec} and three-term~\eqref{MRPdec} energy decompositions make no use of the virial theorem, and can therefore be considered as genuine mass decompositions.

\subsection{Generalized mass decomposition}

So far we insisted on the fact that each term appearing in a genuine mass decomposition should have the meaning of a contribution to proper energy. If we relax this requirement and simply demand that 1) the various terms have the dimension of energy and 2) the sum of the corresponding expectation values gives the mass of the system, then we are led to the concept of \emph{generalized} mass decomposition, which allows one to treat within the same framework various decompositions proposed in the literature. For ease of presentation we will consider in this section only a decomposition of the EMT into quark and gluon contributions, so that each term of the corresponding generalized mass decomposition has the same expression in terms of form factors. It should however be kept in mind that further refinements similar to the ones leading to the three and four-term energy decompositions discussed in Section~\ref{sec:properenergy} can naturally  be considered.

For convenience, let us work in the nucleon rest frame where the total energy coincides with mass. We have already seen that once we have defined a decomposition of the EMT $T^{\mu\nu}=\sum_aT^{\mu\nu}_a$, a natural mass decomposition follows automatically by considering the matrix elements of $T^{00}$ in the rest frame,
\begin{equation}\label{massdec}
    M=\sum_a\frac{\langle p_\text{rest}|\int\ud^3x\,T^{00}_a(x)|p_\text{rest}\rangle}{\langle p_\text{rest}|p_\text{rest}\rangle}=\sum_a\left[A_a(0)+\bar C_a(0)\right]M.
\end{equation}
Combining this rest-frame energy decomposition with the virial theorem
\begin{equation}\label{pressdec}
    0=\sum_a\frac{\langle p_\text{rest}|\int\ud^3x\,T^{ij}_a(x)|p_\text{rest}\rangle}{\langle p_\text{rest}|p_\text{rest}\rangle}=g^{ij}\sum_a\bar C_a(0)\,M,
\end{equation}
and the fact that discrete spacetime symmetries imply
\begin{equation}
    0=\frac{\langle p_\text{rest}|\int\ud^3x\,T^{0i}_a(x)|p_\text{rest}\rangle}{\langle p_\text{rest}|p_\text{rest}\rangle}=\frac{\langle p_\text{rest}|\int\ud^3x\,T^{i0}_a(x)|p_\text{rest}\rangle}{\langle p_\text{rest}|p_\text{rest}\rangle},
\end{equation}
we can define a generalized mass decomposition as follows,
\begin{equation}\label{genSRrest}
    M=\sum_a\frac{\langle p_\text{rest}|\int\ud^3x\,c_{\mu\nu}T^{\mu\nu}_a(x)|p_\text{rest}\rangle}{\langle p_\text{rest}|p_\text{rest}\rangle}=\sum_a\left[A_a(0)+c^\mu_{\phantom{\mu}\mu}\bar C_a(0)\right]M,
\end{equation}
where the $c_{\mu\nu}$ are arbitrary coefficients with the constraint $c_{00}=1$. Some notable examples are:
\begin{itemize}
    \item \emph{Energy decomposition}: $c_{\mu\nu}=g_{0\mu}g_{0\nu}$;
    \item \emph{Trace decomposition}: $c_{\mu\nu}=g_{\mu\nu}$;
    \item \emph{Enthalpy decomposition}: $c_{\mu\nu}=(4g_{0\mu} g_{0\nu}-g_{\mu\nu})/3$;
    \item \emph{Tolman mass decomposition}: $c_{\mu\nu}=2g_{0\mu}g_{0\nu}-g_{\mu\nu}$;
    \item \emph{Light-front momentum decomposition}: $c_{\mu\nu}=2g_{-\mu}g_{-\nu}$;
    \item \emph{Light-front energy decomposition}: $c_{\mu\nu}=2g_{-\mu}g_{+\nu}$.
\end{itemize}
Defining mass as the rest-frame energy naturally leads to the energy decompositions discussed in Refs.~\cite{Ji:1994av,Ji:1995sv,Lorce:2017xzd,Rodini:2020pis,Metz:2020vxd}. Mass being by definition a Lorentz scalar, some authors prefer to relate it to the trace of the EMT, see e.g. Refs.~\cite{Shifman:1978zn,Donoghue:1987av,Kharzeev:1995ij,Hatta:2019lxo,Tanaka:2018nae,Roberts:2021nhw}. Since it is the enthalpy that forms together with the three-momentum a Lorentz four-vector in relativistic thermodynamics~\cite{Pauli:1921,Staruszkiewicz:1966,Moller:1968}, one could also argue that mass is the proportionality factor between four-momentum and four-velocity and hence consider an enthalpy decomposition instead of an energy decomposition. In the context of general relativity, defining the total mass of a system becomes an even more delicate problem owing to contributions associated with the gravitational field. Tolman mass is one of the standard notions of quasi-local mass commonly used because it has ``the great advantage that it can be evaluated by integrating over the region occupied by matter or electromagnetic energy''~\cite{Tolman:1930zz,Tolman:1934,Landau:1951}. In the context of high-energy scatterings, it is particularly convenient to switch to light-front components defined as $a^\pm=(a^0\pm a^3)/\sqrt{2}$, where the $z$-direction is the collision axis. In this formulation of relativistic dynamics, the little group is Galilean and the longitudinal light-front momentum plays in the $(x,y)$-plane the same role as mass does in the non-relativistic context~\cite{Burkardt:2002hr,Abidin:2008sb,Lorce:2018egm,Freese:2021czn}. Finally, coming back to the fact that mass can be seen as the rest-frame energy and noting that light-front boosts are kinematical transformations, one may consider alternatively the light-front version of the energy decomposition discussed e.g. in Refs.~\cite{Lorce:2018egm,Ji:2020baz}.

We note that the generalized mass decomposition introduced in Eq.~\eqref{genSRrest} can easily be expressed in a Lorentz-invariant way according to
\begin{equation}\label{genSRp}
    M=\sum_a\frac{\langle p|\int\ud V\,c_{\mu\nu}T^{\mu\nu}_a(x)|p\rangle}{\langle p|p\rangle}=\sum_a\left[A_a(0)+c^\mu_{\phantom{\mu}\mu}\bar C_a(0)\right]M,
\end{equation}
where $\ud V=u^0\ud^3 x$ is again the invariant proper volume element and the condition on the coefficients is now $c_{\mu\nu}u^\mu u^\nu=1$. For the notable examples presented above, it suffices to apply the substitutions $g_{0\mu}\mapsto u_\mu$ and $g_{\pm\mu}\mapsto n_\pm/(n_\pm\cdot u)\sqrt{2}$ with $n_\pm=(1,\pm \uvec u/|\uvec u|)$. This shows that Poincar\'e symmetry alone does not lead to a unique generalized mass decomposition.

To sum up, while the genuine mass decomposition relies \emph{only} on four-momentum conservation
\begin{equation}
     \sum_a\frac{\langle p|\int\ud^3x\,T^{0\mu}_a(x)|p\rangle}{\langle p|p\rangle}=p^\mu,
\end{equation}
a generalized mass decomposition requires the more general relation
\begin{equation}
     \sum_a\frac{\langle p|\int\ud^3x\,T^{\mu\nu}_a(x)|p\rangle}{\langle p|p\rangle}=\frac{p^\mu p^\nu }{p^0}
\end{equation}
which includes the additional information of mechanical equilibrium expressed by the virial theorem~\eqref{covQFTvirialexpr}. The proper physical interpretation of the contribution associated with subsystem $a$ should therefore account for pressure effects when $c^\mu_{\phantom{\mu}\mu}\neq 1$~\cite{Lorce:2017xzd}.

\section{Operator structure of the energy decomposition}\label{sec:renormalization}

Mass decompositions are often motivated by their operator structures. Some controversy arose recently concerning the form of the renormalized EMT operator in QCD. We discuss in this section this important point, revisiting the structure of the operators in the MS and $\overline{\text{MS}}$  schemes (hereafter, referred as MS-like schemes) in the framework of (conventional)  DR. For a concise description of conventional DR  as well as of other DR procedures we refer to Appendix~\ref{App:DR}, which is based on the original work of Refs.~\cite{Gnendiger:2017pys,Wilson:1972cf,Collins:1984xc}.
We then review the operator structure of the different mass terms in the four-term energy decomposition proposed in Ref.~\cite{Ji:1995sv} and recently revised in Refs.~\cite{Rodini:2020pis,Metz:2020vxd,Ji:2021qgo}. Finally, we discuss the status of the so-called ``quantum anomalous energy'' put forward in the recent works~\cite{Ji:2021mtz,Ji:2021pys,Ji:2021qgo}.

\subsection{Renormalized QCD energy-momentum tensor}
\label{sect:renormalization} 

The EMT is a sum of composite operators, i.e., products of fields and their derivatives evaluated at a single spacetime point. Composite operators are usually divergent in perturbation theory even after Lagrangian renormalization, and require therefore additional renormalization. MS-like renormalization schemes with DR appear to be particularly convenient, since it has been shown that almost all usual algebraic manipulations done at the level of unrenormalized operators remain valid in terms of the renormalized ones~\cite{Breitenlohner:1977hr,Collins:1984xc}. Moreover, both Poincar\'e and gauge symmetries remain exact\footnote{One might naively think that in DR the four-dimensional spacetime is reduced to a $d$-dimensional one with $d<4$, and hence that the original $SO(1,3)$ symmetry is lost. In fact, in order to define non-integer dimensions, spacetime has to be extended to an infinite-dimensional space; see Appendix~\ref{App:DR}. The $SO(1,3)$ symmetry is therefore preserved as a subgroup of a larger symmetry group.} in the intermediate steps. In the following, renormalized operators will be distinguished from unrenormalized ones by a label $R$.  

Nielsen~\cite{Nielsen:1977sy} showed long ago that the total EMT in QCD is finite and does not require additional renormalization, so that $(T^{\mu\nu})_R=T^{\mu\nu}$. Since the individual terms appearing in $T^{\mu\nu}$ do not depend on the spacetime dimension $d$, it follows from the linearity property of MS-like renormalization schemes that~\cite{Suzuki:2013gza,Makino:2014taa,Hatta:2018sqd,Tanaka:2018nae}
 \begin{equation}\label{Tren}
(T^{\mu\nu})_R=(\overline\psi\gamma^{\{\mu}\tfrac{i}{2}\overset{\leftrightarrow}{D}\!\!\!\!\!\phantom{D}^{\nu\}}\psi)_R-(F^{\mu\lambda}F^\nu_{\phantom{\nu}\lambda})_R+\frac{1}{4}\,g^{\mu\nu}(F^2)_R.
\end{equation}
To keep the presentation simple, we omitted terms proportional to the EOM and the gauge non-invariant ones since they do not contribute to the physical matrix elements. Note also that the vacuum expectation values are always implicitly subtracted.

The trace of the QCD EMT is also finite and takes the form~\cite{Collins:1976yq,Nielsen:1977sy,Tarrach:1981bi} 
\begin{equation}\label{Tanom}
    g_{\mu\nu}T^{\mu\nu}=g_{\mu\nu}(T^{\mu\nu})_R=\frac{\beta}{2g}\,(F^2)_R+(1+\gamma_m)(\overline\psi m\psi)_R.
\end{equation}
It differs from the classical trace $g_{\mu\nu}T^{\mu\nu}_\text{class}=\overline{\psi}m\psi=(\overline\psi m\psi)_R$ by a term which is renormalization-group invariant and called the \emph{trace anomaly} 
\begin{equation}
    g_{\mu\nu}(T^{\mu\nu})_R-(\overline{\psi}m\psi)_R=\frac{\beta}{2g}\,(F^2)_R+\gamma_m(\overline\psi m\psi)_R.
    \label{HiddenAnomalyAsEpsF2}
\end{equation}

Looking at the structure of the renormalized EMT~\eqref{Tren}, it is natural to define the renormalized quark and gluon contributions as
\begin{equation}\label{Tqgdef}
    \begin{aligned}
    (T^{\mu\nu}_q)_R&=(\overline\psi\gamma^{\{\mu}\tfrac{i}{2}\overset{\leftrightarrow}{D}\!\!\!\!\!\phantom{D}^{\nu\}}\psi)_R,\\
    (T^{\mu\nu}_g)_R&=-(F^{\mu\lambda}F^\nu_{\phantom{\nu}\lambda})_R+\frac{1}{4}\,g^{\mu\nu}(F^2)_R.
    \end{aligned}
\end{equation}
Since the trace of the QCD EMT is given by
\begin{equation}
    g_{\mu\nu}(T^{\mu\nu})_R=g_{\mu\nu}(T^{\mu\nu}_q)_R+g_{\mu\nu}(T^{\mu\nu}_g)_R,
\end{equation}
we can write\footnote{One can in principle add a term $\propto(\overline\psi \tfrac{i}{2}\overset{\leftrightarrow}{\slashed{D}}\psi)_R$ on the right-hand side of both equations. It is however irrelevant because of the EOM.}, following the notation of Refs.~\cite{Hatta:2018sqd,Tanaka:2018nae},
\begin{equation}\label{xyparam}
    \begin{aligned}
g_{\mu\nu}(T^{\mu\nu}_q)_R&=x\,(F^2)_R+(1+y)(\overline\psi m\psi)_R,\\
g_{\mu\nu}(T^{\mu\nu}_g)_R&=\left(\frac{\beta}{2g}-x\right)(F^2)_R+(\gamma_m-y)(\overline\psi m\psi)_R,
    \end{aligned}
\end{equation}
where $x$ and $y$ are finite numbers of order $\ucal O(\alpha_s)$ which parametrize how the anomalous contributions to the trace are shared between the quark and gluon parts of the EMT. 

\subsection{Operator mixing}

We sketch here the construction of the renormalized EMT operators in the  MS-like schemes with DR as defined in Appendix~\ref{App:DR}, and we refer to~\cite{Suzuki:2013gza,Makino:2014taa,Hatta:2018sqd,Tanaka:2018nae,Metz:2020vxd} for more details. Renormalization through normal products (defining the finite part of composite operators) and the emergence of the anomaly in DR have both been studied long ago in Refs.~\cite{Breitenlohner:1977hr,Bonneau:1980zp,Bonneau:1979jx,Collins:1976yq,Nielsen:1977sy,Tarrach:1981bi,Collins:1984xc}. An explicit application to the $O(N)$ nonlinear sigma model in $\overline{\text{MS}}$ scheme is given 
in Ref.~\cite{Caracciolo:2000td}.

The renormalized operators are obtained from a basis of bare composite operator $O_i$ as follows,
\begin{equation}
    (O_i)_R=\sum_jZ_{ij}O_j.\label{eq:ren}
\end{equation}
As discussed in Refs.~\cite{Makino:2014taa,Tanaka:2018nae,Hatta:2019lxo,Hatta:2018sqd,Metz:2020vxd,Rodini:2020pis}, the remormalization of the QCD EMT involves four independent operators $O_i$ that mix through  Eq.~\eqref{eq:ren} via 10 renormalization constants. 
Explicitly, one has the following system of equations,
\begin{equation}\label{general_renorm}
    \begin{pmatrix}
    (\overline\psi\gamma^{\{\mu}\tfrac{i}{2}\overset{\leftrightarrow}{D}\!\!\!\!\!\phantom{D}^{\nu\}}\psi)_R\\
    -(F^{\mu\lambda}F^\nu_{\phantom{\nu}\lambda})_R\\
    g^{\mu\nu}(F^2)_R\\
    g^{\mu\nu}(\overline\psi m\psi)_R
    \end{pmatrix}=
    \begin{pmatrix}
    Z_{qq}&Z_{qg}&Z_{qF}&Z_{qm}\\
    Z_{gq}&Z_{gg}&Z_{gF}&Z_{gm}\\
    0&0&Z_{FF}&Z_{Fm}\\
    0&0&0&1
    \end{pmatrix}
     \begin{pmatrix}
     \overline\psi\gamma^{\{\mu}\tfrac{i}{2}\overset{\leftrightarrow}{D}\!\!\!\!\!\phantom{D}^{\nu\}}\psi\\
     -F^{\mu\lambda}F^\nu_{\phantom{\nu}\lambda}\\
    g^{\mu\nu}F^2\\
    g^{\mu\nu}\overline\psi m\psi
    \end{pmatrix},
\end{equation}
where the operators with vanishing contribution to the physical matrix elements have consistently been omitted.

Thanks to Lorentz symmetry, one can alternatively regroup the operators to form scalar and symmetric traceless tensor representations of the Lorentz group which do not mix under renormalization. This amounts to changing the operator basis in such a way that the renormalization matrix in Eq.~\eqref{general_renorm} turns into a block-diagonal form, i.e.,
\begin{equation}\label{twist2}
    \begin{pmatrix}
    (\overline\psi\gamma^{\{\mu}\tfrac{i}{2}\overset{\leftrightarrow}{D}\!\!\!\!\!\phantom{D}^{\nu\}}\psi)_R- \frac{g^{\mu\nu}}{d}\,g_{\alpha\beta}(\overline\psi\gamma^{\{\alpha}\tfrac{i}{2}\overset{\leftrightarrow}{D}\!\!\!\!\!\phantom{D}^{\beta\}}\psi)_R\\
    -(F^{\mu\lambda}F^\nu_{\phantom{\nu}\lambda})_R+\frac{g^{\mu\nu}}{d}\,g_{\alpha\beta}(F^{\alpha\lambda}F^\beta_{\phantom{\beta}\lambda})_R\\
      g^{\mu\nu}(F^2)_R\\
    g^{\mu\nu}(\overline\psi m\psi)_R
    \end{pmatrix}=
    \begin{pmatrix}
    Z_{qq}&Z_{qg}&0&0\\
    Z_{gq}&Z_{gg}&0&0\\
    0&0&Z_{FF}&Z_{Fm}\\
    0&0&0&1
    \end{pmatrix}
     \begin{pmatrix}
     \overline\psi\gamma^{\{\mu}\tfrac{i}{2}\overset{\leftrightarrow}{D}\!\!\!\!\!\phantom{D}^{\nu\}}\psi-\frac{g^{\mu\nu}}{d}\,g_{\alpha\beta}\overline\psi\gamma^{\{\alpha}\tfrac{i}{2}\overset{\leftrightarrow}{D}\!\!\!\!\!\phantom{D}^{\beta\}}\psi\\
     -F^{\mu\lambda}F^\nu_{\phantom{\nu}\lambda}+\frac{g^{\mu\nu}}{d}\,g_{\alpha\beta}F^{\alpha\lambda}F^\beta_{\phantom{\beta}\lambda}\\
    g^{\mu\nu}F^2\\
    g^{\mu\nu}\overline\psi m\psi
    \end{pmatrix}.
\end{equation}
By construction the renormalized operators remain finite in the limit $d\to 4$. As a result, we can safely express the traceless operators as on the LHS of Eq.~\eqref{twist2} with $d$ replaced by $4$. Because of the linearity property of the renormalization~\eqref{eq:ren}, the two procedures in Eqs.~\eqref{general_renorm} and \eqref{twist2} are perfectly equivalent. There is therefore in practice no distinction between what the authors of Ref.~\cite{Ji:2021qgo} call the ``standard'' and ``non-standard'' way of renormalizing operators. The only crucial point is that one has to be careful with the way of writing properly the renormalized traceless operators, an aspect that will be discussed in more detail in Section~\ref{sec:tracelessop}.

The EMT renormalization constants in the MS-like schemes
have been derived up to two and three loops in Refs.~\cite{Tanaka:2018nae,Hatta:2019lxo}, and further discussed in Refs.~\cite{Hatta:2018sqd,Metz:2020vxd,Rodini:2020pis} in the context of various mass sum rules. 
 In these schemes, using  DR and $d=4-2\epsilon$,
the structure  of the renormalization constants is
\begin{eqnarray}
Z_{ij}\big|_{\mathrm{MS}} &=& \delta_{ij}+ \alpha_s \, \frac{a_{ij,1}}{\epsilon} + \alpha_s^2 \, \bigg( \frac{b_{ij,2}}{\epsilon^2} + \frac{b_{ij,1}}{\epsilon} \bigg) + \alpha_s^3 \, \bigg( \frac{c_{ij,3}}{\epsilon^3} + \frac{c_{ij,2}}{\epsilon^2} + \frac{c_{ij,1}}{\epsilon} \bigg) \,, 
\label{Z_ms_def}\\
Z_{ij} \big|_{\overline{\text{MS}}} &=& \delta_{ij} + \alpha_s \, \frac{\bar a_{ij,1}}{\epsilon} \, S_{\epsilon} + \alpha_s^2 \, \bigg( \frac{\bar b_{ij,2}}{\epsilon^2} + \frac{\bar b_{ij,1}}{\epsilon} \bigg) \, S^2_{\epsilon} + \alpha_s^3 \, \bigg( \frac{\bar c_{ij,3}}{\epsilon^3} + \frac{\bar c_{ij,2}}{\epsilon^2} + \frac{\bar c_{ij,1}}{\epsilon} \bigg) \, S^3_{\epsilon} \,,
\label{Z_msbar_def}
\end{eqnarray}
where the finite quantity $S_{\epsilon}$ can follow different conventions~\cite{Collins:2011zzd,Bardeen:1978yd}, with an expansion in powers of $\epsilon$ that differs at ${\cal O}(\epsilon^2)$ and higher.  We refrain from providing here the explicit form of the  renormalization factors, but they can be found in Refs.~\cite{Hatta:2018sqd,Tanaka:2018nae}, and also in~\cite{Metz:2020vxd} in different $\overline{\text{MS}}$ schemes. We notice that, formally, the MS renormalization factors can be obtained from the $\overline{\text{MS}}$ ones by simply  setting $S_\epsilon = 1$.

When dealing with tensor operators, one has to pay particular attention to the manipulation of the trace and renormalization operations since, in general, they do not commute~\cite{Collins:1984xc,Bonneau:1979jx},
\begin{equation}\label{noncomm}
g_{\mu\nu}(O^{\mu\nu})_R\neq (g_{\mu\nu}O^{\mu\nu})_R.
\end{equation}
In DR, this arises from the fact that the trace operation may change the pole structure and hence the result of the normal product. The non-commutativity of these operations is a reflection of the trace anomaly. In some other renormalization schemes, like for instance BPHZ, these operations do commute but linearity is lost~\cite{Collins:1974da,Collins:1984xc}. The general message is that because of the trace anomaly it is impossible to preserve all the algebraic manipulations under renormalization.
To clarify this point, we consider an explicit example. As outlined in Appendix~\ref{App:DR}, any DR scheme is well-defined only in perturbation theory and consists of replacing  the four-dimensional loop integration with the map $\Id$, and mapping all vectors from the (eventually Wick rotated) four-dimensional Minkowski space into the infinitely-dimensional QS$_d$ space. We remind that relations among operators are usually understood as relations valid for the corresponding Green functions.
Indeed, the relation 
\[
g_{\mu\nu} \overline\psi iD^\mu \gamma^\nu\psi = \overline\psi i\slashed{D}\psi
\]
has to be understood at the level of the matrix elements
\[
g_{\mu\nu} \langle\overline\psi iD^\mu \gamma^\nu\psi\rangle = \langle\overline\psi i\slashed{D}\psi\rangle.
\]
For the bare operator, one has
\[
g_{\mu\nu}T_q^{\mu\nu} = \overline\psi \tfrac{i}{2}\overset{\leftrightarrow}{\slashed{D}}\psi = \overline\psi m\psi.
\]
For the renormalized operator, instead, one obtains, using Eq.~\eqref{general_renorm},
\[
g_{\mu\nu}(T^{\mu\nu}_q)_R = (Z_{qq} +d\,Z_{qm})\, \overline\psi m\psi + (-Z_{qg}+d\,Z_{qF})\,F^2,
\label{QuarkRenormTrace}
\]
which can be written as 
\[
g_{\mu\nu}(T^{\mu\nu}_q)_R = (1+c_1)\overline\psi m\psi + c_2 F^2 = g_{\mu\nu}T^{\mu\nu}_q + c_1 \overline\psi m\psi + c_2 F^2 ,
\]
where both $c_1$ and $c_2$ start at $\ucal O(\alpha_s)$ in perturbation theory and are defined as
\[
c_1 = Z_{qq}-1+d\,Z_{qm}, \quad c_2 =-Z_{qg}+d\,Z_{qF}.
\]
The \textit{total} contribution $c_1 \overline\psi m\psi + c_2 F^2$ can be interpreted as a finite correction to the  trace of the bare operator\footnote{
A very similar example (with explicit one-loop calculations) is given in Chapter 6.5 of Ref. \cite{Collins:1984xc}.}.

\subsection{Symmetric traceless operators}\label{sec:tracelessop}

It appears that there is some confusion in the literature about the form of the symmetric traceless operators. One reason is that many textbooks and papers do not spell out explicitly the trace terms and simply write
\begin{equation}
    \bar O^{\mu\nu}\equiv O^{\{\mu\nu\}}-\text{trace}.
\end{equation}
This lack of explicitness can be understood from the observation that, in the context of high-energy scattering, one is often only interested in the specific light-front component $T^{++}_a$, representing the light-front density of longitudinal momentum. Since $g^{++}=0$, one does not need to worry in this case about the explicit form of the trace terms. The problem arises however as soon as one considers components with a non-vanishing contribution from the metric, like e.g.~the energy density $T^{00}_a$.

The unambiguous definition of the symmetric traceless part of a generic rank-two tensor operator in $d$-dimensional spacetime is
\begin{equation}\label{deftraceless}
    \bar O^{\mu\nu}\equiv O^{\{\mu\nu\}}-\frac{1}{d}\,g^{\mu\nu}g_{\alpha\beta}O^{\alpha\beta}.
\end{equation}
In particular, for a renormalized operator the explicit expression is~\cite{Wilson:1970wp}
\begin{equation}\label{deftracelessR}
    (\bar O^{\mu\nu})_R\equiv(O^{\{\mu\nu\}})_R-\frac{1}{d}\,g^{\mu\nu}g_{\alpha\beta}(O^{\alpha\beta})_R.
\end{equation}
This tensor is manifestly traceless irrespective of whether the trace and renormalization operations commute or not. Also, since the explicit $d$-dependence appears outside of the normal product, we can safely replace $d$ by $4$. 

Renormalization should preserve Lorentz symmetry, and so operators belonging to different representations of the Lorentz group should not mix with each other. We can therefore write~\cite{Peskin:1995ev}
\begin{equation}
    (\bar O^{\mu\nu}_i)_R=\sum_jZ_{ij}\bar O^{\mu\nu}_j\label{eq:rentraceless}
\end{equation}
which is obviously compatible by linearity with Eqs.~\eqref{eq:ren},~\eqref{deftraceless} and~\eqref{deftracelessR}. The standard shorthand notation $(\bar O^{\mu\nu}_i)_R$ is however somewhat misleading, because it gives an incentive to write
\begin{equation}\label{misleadingexp}
    (\bar O^{\mu\nu})_R= \left(O^{\{\mu\nu\}}-\frac{1}{d}\,g^{\mu\nu}g_{\alpha\beta}O^{\alpha\beta}\right)_R,
\end{equation}
an expression which must be treated with great care. For example, while renormalized operators are by construction finite in the limit $d\to 4$, one should in general refrain from replacing directly $d$ by $4$ in Eq.~\eqref{misleadingexp}. Indeed, in MS-like schemes with DR the notation $(O)_R$ means that one removes the contributions of the operator $O$ which diverge as $d\to 4$. The latter limit must then be considered at the very end of a calculation and cannot be applied directly to the expression inside the brackets in Eq.~\eqref{misleadingexp}. For the same reason, one must also pay attention that in general
\begin{equation}
     (\bar O^{\mu\nu})_R\neq (O^{\{\mu\nu\}})_R-\frac{1}{d}\,g^{\mu\nu}(g_{\alpha\beta}O^{\alpha\beta})_R
\end{equation}
because of Eq.~\eqref{noncomm}.

As a result of the above discussion, the renormalized symmetric traceless quark and gluon operators are unambiguously given by~\cite{Suzuki:2013gza,Makino:2014taa,Hatta:2018sqd,Tanaka:2018nae}
\begin{equation}
    \begin{aligned}\label{qgtraceless1}
        (\bar T^{\mu\nu}_q)_R&= (\overline\psi\gamma^{\{\mu}\tfrac{i}{2}\overset{\leftrightarrow}{D}\!\!\!\!\!\phantom{D}^{\nu\}}\psi)_R- \frac{g^{\mu\nu}}{4}\,g_{\alpha\beta}(\overline\psi\gamma^{\{\alpha}\tfrac{i}{2}\overset{\leftrightarrow}{D}\!\!\!\!\!\phantom{D}^{\beta\}}\psi)_R,\\
        (\bar T^{\mu\nu}_g)_R&= -(F^{\mu\lambda}F^\nu_{\phantom{\nu}\lambda})_R+\frac{g^{\mu\nu}}{4}\,g_{\alpha\beta}(F^{\alpha\lambda}F^\beta_{\phantom{\beta}\lambda})_R,\\
    \end{aligned}
\end{equation}
and can alternatively be expressed as
\begin{equation}\label{qgtraceless2}
    \begin{aligned}
    (\bar T^{\mu\nu}_q)_R&=(\overline\psi\gamma^{\{\mu}\tfrac{i}{2}\overset{\leftrightarrow}{D}\!\!\!\!\!\phantom{D}^{\nu\}}\psi)_R-\frac{g^{\mu\nu}}{4}\left[x\,(F^2)_R+(1+y)(\overline\psi m\psi)_R\right],\\
    (\bar T^{\mu\nu}_g)_R&=-(F^{\mu\lambda}F^\nu_{\phantom{\nu}\lambda})_R+\frac{g^{\mu\nu}}{4}\left[\left(1+x-\frac{\beta}{2g}\right)(F^2)_R+(y-\gamma_m)(\overline\psi m\psi)_R\right],
    \end{aligned}
\end{equation}
using Eq.~\eqref{xyparam}. One could also formally write
\begin{equation}\label{qgtraceless3}
    \begin{aligned}
    (\bar T^{\mu\nu}_q)_R&=\lim_{d\to 4}\left(\overline\psi\gamma^{\{\mu}\tfrac{i}{2}\overset{\leftrightarrow}{D}\!\!\!\!\!\phantom{D}^{\nu\}}\psi-\frac{g^{\mu\nu}}{d}\,\overline\psi m\psi\right)_R,\\
    (\bar T^{\mu\nu}_g)_R&=\lim_{d\to 4}\left(-F^{\mu\lambda}F^\nu_{\phantom{\nu}\lambda}+\frac{g^{\mu\nu}}{d}\,F^2\right)_R.
    \end{aligned}
\end{equation}
Once again, it is essential that the limit $d\to 4$ is taken \emph{after} minimal subtraction.

Based on their explicit operator expressions in MS-like scheme with DR, we conclude that there is in general no simple physical interpretation for $(\bar T^{00}_q)_R$ or $(\bar T^{00}_g)_R$. We discuss in the following the consequences for the energy decomposition.

\subsection{Energy decomposition}\label{sec:energydecren}

Following the approach of the original works on the nucleon mass decomposition~\cite{Ji:1994av,Ji:1995sv}, the total renormalized EMT can be obtained by adding the renormalized traceless and trace parts
\begin{equation}
    (T^{\mu\nu})_R=(\bar T^{\mu\nu})_R+\frac{g^{\mu\nu}}{4}\,g_{\alpha\beta}(T^{\alpha\beta})_R.
\end{equation}
Using the incentive form for the renormalized traceless operators given in Eq.~\eqref{qgtraceless3}, we get explicitly
\begin{equation}\label{TJicorr}
\begin{aligned}
    (T^{\mu\nu})_R&=\lim_{d\to 4}\left(\overline\psi\gamma^{\{\mu}\tfrac{i}{2}\overset{\leftrightarrow}{D}\!\!\!\!\!\phantom{D}^{\nu\}}\psi-\frac{g^{\mu\nu}}{d}\,\overline\psi m\psi\right)_R+\lim_{d\to 4}\left(-F^{\mu\lambda}F^\nu_{\phantom{\nu}\lambda}+\frac{g^{\mu\nu}}{d}\,F^2\right)_R\\
    &\quad +\frac{1}{4}\,g^{\mu\nu}\left[\frac{\beta}{2g}\,(F^2)_R+(1+\gamma_m)(\overline\psi m\psi)_R\right].
    \end{aligned}
\end{equation}
In order to obtain a decomposition of energy at the operator level, we consider $(T^{00})_R$ and integrate over space.
Following the original derivation~\cite{Ji:1994av,Ji:1995sv},
 the QCD Hamiltonian 
\begin{equation}
     H=H_T+H_S\label{eq:sumrule}
\end{equation}
has been decomposed into a traceless (tensor) 
and trace (scalar) part as~\cite{Ji:2021mtz}
\begin{equation}\label{eq:sumrule2}
\begin{aligned}
H_T&\equiv\int\ud^3x\,(\bar T^{00})_R=(H_q+ H_g)+\frac{3}{4}\, H_m,\\
H_S&\equiv\frac{1}{4}\int\ud^3x\,g_{\alpha\beta}(T^{\alpha\beta})_R=H_\text{a}+\frac{1}{4}\,H_m,\
\end{aligned}
\end{equation}
where
\begin{equation}\label{new-energy}
\begin{aligned}
   H_\text{a}&=\frac{1}{4}\int\ud^3x\left[\frac{\beta}{2g}\,(F^2)_R+\gamma_m (\overline\psi m\psi )_R\right],\\
H_m &= \int\ud^3x\,(\overline{\psi}m \psi)_R,\\
 H_q+ H_g&=\lim_{\epsilon\to 0}\int\ud^3x\,\left(\psi^\dag\, \uvec{\alpha}\cdot i\uvec{D}\,\psi+\frac{2-2\epsilon}{4-2\epsilon}\,\uvec E^2+\frac{2}{4-2\epsilon}\,\uvec B^2 \right)_R
\end{aligned}
\end{equation}
are three contributions that are separately renormalization group invariant. The expansion of the last contribution in powers of $\epsilon$ gives 
\begin{equation}
   H_q+ H_g=\int\ud^3x \left\{(\psi^\dagger \,\uvec{\alpha}\cdot  i \uvec{D}  \, \psi)_R+\frac{(\uvec E^2+\uvec B^2)_R}{2}-\frac{1}{4}\left[\frac{\beta}{2g}\,(F^2)_R+\gamma_m(\overline\psi m\psi)_R\right]\right\},
\end{equation}
using the linearity property of MS-like scheme and the trace anomaly relation~\cite{Collins:1976yq,Nielsen:1977sy,Tarrach:1981bi}
\begin{equation}
    \epsilon(\uvec E^2-\uvec B^2)=-\frac{\epsilon}{2}\,F^2=\frac{\beta}{2g}\,(F^2)_R+\gamma_m(\overline\psi m\psi)_R.
\end{equation}

In conclusion, $(H_q+H_g)$ contains an anomalous contribution which compensates exactly $H_\text{a}$ in Eq.~\eqref{eq:sumrule2}. No anomalous contribution survives therefore in the energy budget, which is then composed of three terms instead of four~\cite{Metz:2020vxd,Rodini:2020pis}
\begin{equation}\label{ThreeTermsSR}
    H= \widetilde{H}_q+ \widetilde{H}_g + H_m, 
\end{equation}
with
\begin{equation}
    \begin{aligned}
       \widetilde H_q&=\int\ud^3x\,(\psi^\dagger \,\uvec{\alpha}\cdot  i \uvec{D}  \, \psi )_R=\int\ud^3x\,(T^{00}_q)_R-H_m,\\
       \widetilde H_g&=\int\ud^3x\,\frac{(\uvec E^2+\uvec B^2)_R}{2}=\int\ud^3x\,(T^{00}_g)_R.
    \end{aligned}
\end{equation}
This structure follows also directly from Eq.~\eqref{Tren} without the need of decomposing first the EMT into traceless and trace parts. Moreover, one can safely interpret $\widetilde H_q$ and $\widetilde H_g$ as the quark and gluon kinetic+potential energies, in agreement with the tensor analysis approach. 

\subsection{Diagonal schemes}

While agreeing formally with the results of the previous section, the authors of Ref.~\cite{Ji:2021qgo} complained that
\begin{equation}
    (T^{00}_g)_R=-(F^{0\lambda}F^0_{\phantom{0}\lambda})_R+\frac{1}{4}\,(F^2)_R=(\uvec E^2)_R+\tfrac1{2}(\uvec B^2-\uvec E^2)_R=\tfrac{1}{2}(\uvec E^2+\uvec B^2)_R
\end{equation}
mixes the tensor and scalar representations of the Lorentz group, and claimed that the notation $\tfrac{1}{2}(\uvec E^2+\uvec B^2)_R$ has commonly been reserved for $(\bar T^{00}_g)_R$ and not $(T^{00}_g)_R$, referring to the works~\cite{Luke:1992tm,Kharzeev:1995ij}. Our opinion is that the latter statement is a misrepresentation of what can be found in the literature. In both papers~\cite{Luke:1992tm,Kharzeev:1995ij} the renormalized traceless gluon operator indeed appears in its classical form, but we observe that neither the corresponding quark operator nor the renormalization scheme are specified. These works are in fact inspired by an old seminal paper of Voloshin and Zakharov~\cite{Voloshin:1980zf}, where it is suggested that one can measure the gluonic part of the trace anomaly using quarkonia. It appears that Voloshin and Zakharov used the relation $g_{\mu\nu}g_{\alpha\beta}(F^{\mu\alpha}F^{\nu\beta})_R=(F^2)_R$ to derive a low-energy theorem. This indicates that they are not working in a MS-like renormalization scheme but in another one where $(\bar T^{\mu\nu}_g)_R=(T^{\mu\nu}_g)_R$, so that the EMT trace (including the anomalous contributions) arises solely from the quark sector $g_{\mu\nu}(T^{\mu\nu})_R=g_{\mu\nu}(T^{\mu\nu}_q)_R$. Other choices have also been made in the literature. For example, in a comment to Ref.~\cite{Voloshin:1980zf}, Novikov and Shifman~\cite{Novikov:1980fa} wrote the total renormalized gluon EMT $(T^{\mu\nu}_g)_R$ in the classical form, in agreement with Eq.~\eqref{Tqgdef} and Refs.~\cite{Novikov:1981xi,Shifman:1988zk}, but they required that the trace is given by $g_{\mu\nu}(T^{\mu\nu}_g)_R=\frac{\beta_F}{2g}(F^2)_R$, where $\beta_F$ is the contribution to the beta function arising from gluon loops only. This indicates that yet a different renormalization scheme has been chosen\footnote{Adding to the confusion, in a recent paper~\cite{Kharzeev:2021qkd} the temporal component of the gluon part of the QCD EMT is denoted $T^{00}_g=\frac{1}{2}(\uvec E^2+\uvec B^2)$ with a reference to the work of Novikov and Shifman~\cite{Novikov:1980fa}, while at the same time it is presented as a $2^{++}$ gluon operator, i.e. a symmetric traceless tensor, like in the work of Voloshin and Zakharov~\cite{Voloshin:1980zf}.}. This example shows the importance of clearly specifying the renormalization scheme, for otherwise a comparison between different works may lead to apparent contradictions.

When discussing the form of the gluon operators and their properties, one must not forget the quark sector. We have seen that in MS-like schemes the traces of the quark and gluon contributions to the EMT~\eqref{xyparam} involve some mixing parametrized by two finite numbers $x$ and $y$. It follows from the unambiguous definition~\eqref{deftracelessR} that the renormalized traceless quark and gluon operators can also be expressed in a way that involves explicitly $x$ and $y$, see Eq.~\eqref{qgtraceless2}. The values of these parameters are directly determined by the renormalization factors. Their explicit expressions in MS-like schemes with DR can be found in Refs.~\cite{Hatta:2018sqd,Tanaka:2018nae,Hatta:2019lxo,Metz:2020vxd}. They are quite cumbersome and indicate that due to operator mixing the trace anomaly is shared in a nontrivial way between the quark and gluon contributions. 

Simpler expressions for the renormalized operators can however be obtained by applying a finite renormalization to the MS-like operators. The only effect of this finite renormalization will be to reshuffle the anomalous contributions between $g_{\mu\nu}(T^{\mu\nu}_q)_R$ and $g_{\mu\nu}(T^{\mu\nu}_g)_R$, i.e., changing the values of $x$ and $y$. The total anomaly remains however unchanged. It is through such a finite renormalization that one can connect in principle the operators in MS-like scheme with DR to the ones discussed in Refs.~\cite{Luke:1992tm,Kharzeev:1995ij,Voloshin:1980zf,Novikov:1980fa}.

The so-called diagonal schemes~\cite{Rodini:2020pis,Metz:2020vxd} keep the mixing between quark and gluon operators under the trace operation as simple as possible. We present here three of the most meaningful choices:
\begin{itemize}
\item \emph{D1 scheme} -- One may choose a scheme where the quark and gluon operators do not mix under the trace operation~\cite{Rodini:2020pis}. It corresponds to the choice $x=0$ and $y=\gamma_m$ so that
\begin{equation}\label{D1tr}
    \begin{aligned}
    g_{\mu\nu}(T^{\mu\nu}_q)_\text{D1}&=(1+\gamma_m)(\overline\psi m\psi)_R,\\
    g_{\mu\nu}(T^{\mu\nu}_g)_\text{D1}&=\frac{\beta}{2g}\,(F^2)_R,
    \end{aligned}
\end{equation}
which was the situation considered in Ref.~\cite{Lorce:2017xzd}, allowing one to identify the quark and gluon contributions to the EMT trace used in Refs.~\cite{Ji:1994av,Ji:1995sv} with the corresponding traces of the quark and gluon contributions to the EMT.
\item \emph{D2 scheme} -- Since the whole anomaly $\big[\frac{\beta}{2g}\,(F^2)_R+\gamma_m(\overline\psi m\psi)_R\big]$ and $(\bar\psi m\psi)_R$ are separately renormalization-group invariant, one may prefer to work with $x=y\,\frac{\beta}{2g\gamma_m}$. The D2 scheme introduced in Ref.~\cite{Metz:2020vxd} corresponds to the choice $x=y=0$ and attributes all the anomalous terms to the renormalized gluon contribution to the EMT,
\begin{equation}\label{D2tr}
    \begin{aligned}
    g_{\mu\nu}(T^{\mu\nu}_q)_\text{D2}&=(\overline\psi m\psi)_R,\\
    g_{\mu\nu}(T^{\mu\nu}_g)_\text{D2}&=\frac{\beta}{2g}\,(F^2)_R+\gamma_m(\overline\psi m\psi)_R.
    \end{aligned}
\end{equation}
\item \emph{D3 scheme} -- For completeness, we mention a third possibility corresponding to the choice $x=\frac{\beta}{2g}$ and $y=\gamma_m$,
\begin{equation}\label{D3tr}
    \begin{aligned}
    g_{\mu\nu}(T^{\mu\nu}_q)_\text{D3}&=\frac{\beta}{2g}\,(F^2)_R+(1+\gamma_m)(\overline\psi m\psi)_R,\\
    g_{\mu\nu}(T^{\mu\nu}_g)_\text{D3}&=0,
    \end{aligned}
\end{equation}
which attributes all the anomalous terms to the renormalized quark contribution to the EMT.
\end{itemize}

\subsection{``Quantum anomalous energy''}

In a series of recent papers~\cite{Ji:2021pys,Ji:2021mtz,Ji:2021qgo}, the concept of ``quantum anomalous energy'' (QAE) has been emphasized as a key aspect of the nucleon mass structure. QAE finds its origin in the four-term energy decomposition proposed in Refs.~\cite{Ji:1994av,Ji:1995sv}, where it has been argued that in MS-like renormalization with DR the QCD Hamiltonian takes the form
\begin{equation}
    H=\int\ud^3x\,\left[(\psi^\dagger \,\uvec{\alpha}\cdot  i \uvec{D}  \, \psi)_R+(\overline\psi m\psi)_R+\frac{(\uvec E^2+\uvec B^2)_R}{2}\right]+\frac{1}{4}\int\ud^3x\,\left[\frac{\beta}{2g}\,(F^2)_R+\gamma_m(\overline\psi m\psi)_R\right].
\end{equation}
It seems therefore that the renormalized QCD Hamiltonian receives on top of its classical form a new contribution equal to a quarter of the trace anomaly. This contribution is unexpected and referred to as QAE.

The analysis of Refs.~\cite{Ji:1994av,Ji:1995sv} has been revisited in Refs.~\cite{Rodini:2020pis,Metz:2020vxd} with the new conclusion that no anomalous contributions actually survive in the Hamiltonian; see our discussion in Section~\ref{sec:energydecren}. The difference between these two analyses can be traced back to the renormalized traceless operators. While those operators did not appear explicitly in the original works~\cite{Ji:1994av,Ji:1995sv}, their precise form used in the context of the four-term decomposition has been specified in a recent work~\cite{Ji:2021mtz}. Contrary to Eq.~\eqref{qgtraceless3}, it appears that the $d\to 4$ limit is taken \emph{before} the normal product, giving the impression that the traceless operators can be written in the same way as in the classical case. As stressed in Section~\ref{sec:tracelessop}, this is an incorrect notation since it assumes that trace and renormalization are commuting operations, a property that is not satisfied in general in MS-like schemes with DR. 

Writing the renormalized QCD EMT as a sum of a traceless part with classical form and a trace part with anomalous contributions like in Refs.~\cite{Ji:1994av,Ji:1995sv,Ji:2021pys,Ji:2021mtz,Ji:2021qgo} may seem a priori attractive, since it gives the impression that the anomaly is distributed equally between the diagonal components of the EMT. As a result, the Hamiltonian $H=\int\ud^3x\,(T^{00})_R$ is expected to provide a quarter of the trace anomaly, considered in these papers as a ``new'' form of energy. This picture is however inconsistent with Poincar\'e symmetry. Indeed, time translation is an \emph{exact} symmetry of the theory. The form of the corresponding generator, i.e. the Hamiltonian, must then be the same as in the classical case as a consequence of the quantum action principle~\cite{Breitenlohner:1977hr,Collins:1984xc}. 

As indicated by its name, the trace anomaly is a pure quantum contribution associated with the trace of the EMT, and not with its individual diagonal components. It expresses the breaking of spacetime dilatations, and not a breaking of spacetime translations. At the operator level, the trace anomaly has nothing to do with the Hamiltonian. Motivated by Lorentz symmetry and deep-inelastic scattering experiments\footnote{We remind that in high-energy scattering experiments, one is mostly sensitive to the light-front component $(T^{++})_R$ of the EMT. One can then use Lorentz symmetry to relate the matrix elements of $(T^{++})_R$ to those of $(\bar T^{\mu\nu})_R$. The fact that the anomalous contributions do not appear in $(T^{++})_R$ as a consequence of $g^{++}=0$ does not however imply that they should also be absent in $(\bar T^{\mu\nu})_R$.}, one may of course decompose the Hamiltonian into tensor and scalar parts as in Eq.~\eqref{eq:sumrule}, but this does not bring much fundamental insight since it just amounts to writing the Hamiltonian as
\begin{equation} \label{hamiltonian_decomp}
    H=(H-H_S)+H_S,
\end{equation}
where $H_S$ contains anomalous contributions (see Eq.~\eqref{eq:sumrule2}) while $H$ does not. As already stressed in Section~\ref{sec:reltracemass}, the only way to relate non-trivially the Hamiltonian, and hence the mass of a system, to the trace anomaly is at the level of matrix elements. Indeed, the virial theorem~\eqref{virialthmprest} tells us that the expectation value of the stress tensor must vanish at rest. As a result, we can write
\begin{equation}
    M=\frac{\langle p_{\text{rest}}|\int\ud^3x\,(T^{00})_R(x)|p_{\text{rest}}\rangle}{\langle p_{\text{rest}}|p_{\text{rest}}\rangle}=\frac{\langle p_{\text{rest}}|\int\ud^3x\,g_{\mu\nu}(T^{\mu\nu})_R(x)|p_{\text{rest}}\rangle}{\langle p_{\text{rest}}|p_{\text{rest}}\rangle},
\end{equation}
which leads to the relation~\cite{Rodini:2020pis,Metz:2020vxd}
\begin{equation}
\langle p_{\text{rest}}|\, ( \psi^{\dagger} \, \bm{\alpha}\cdot i \bm{D}  \, \psi )_R + \frac{1}{2}\,(\uvec E^2 +\uvec B^2)_R \,|p_{\text{rest}} \rangle
= \langle p_{\text{rest}}|\, \gamma_m (  \overline{\psi}m \psi )_R + \frac{\beta}{2g}\, ( F^2 )_R \,|p_{\text{rest}} \rangle,
\label{me_equivalence}
\end{equation}
It is then clear that one can have a mass sum rule with contributions from \emph{either} the parton energies \emph{or} the anomaly, but a sum rule with both contributions at the same time does not appear naturally.

In summary, the renormalized Hamiltonian in QCD does not contain anomalous contributions since it is protected by translation symmetry. The so-called QAE given by the expectation value of $H_\text{a}$, defined as a quarter of the trace anomaly and appearing in the ``scalar'' part of the Hamiltonian, does not provide clear fundamental insight since it is exactly compensated by the same contribution with opposite sign from the ``tensor'' part of the Hamiltonian. The latter does not appear in Refs.~\cite{Ji:1994av,Ji:1995sv,Ji:2021pys,Ji:2021mtz,Ji:2021qgo} due to an unjustified notation for the renormalized traceless operators in MS-like scheme with DR.

\section{EMT decomposition on the lattice}\label{sec:lattice}

Recent papers~\cite{Ji:2021pys,Ji:2021mtz,Ji:2021qgo} try to justify the appearance of QAE in the nucleon mass budget based on some works by Rothe~\cite{Rothe:1995hu,Rothe:1995av} in the context of lattice QCD (LQCD). It appears that the question of the Hamiltonian in LQCD is an old and difficult problem. Contrary to DR, lattice regularization allows one to renormalize the theory in a non-perturbative way. On the other hand, Poincar\'e symmetry is broken by the introduction of a finite lattice spacing. One must therefore pay particular attention that in the limit of vanishing lattice spacing the Poincar\'e symmetry is correctly recovered. It also means that one has to be careful with the physical interpretation of lattice expressions, since the breaking of Poincar\'e symmetry by the regulator generates artifacts, especially in currents associated with spacetime symmetries like the EMT. Unfortunately, this essential aspect of the problem has not been considered in Refs.~\cite{Ji:2021pys,Ji:2021mtz,Ji:2021qgo}. We show in the following that taking it into account sheds light on the results presented in these papers, and leads to the conclusion that the relevant LQCD papers actually do not provide concrete support for the concept of QAE.

\subsection{Lattice sum rules}

Using the Wilson action~\cite{Wilson:1974sk}, Michael found that the glueball mass can be expressed as~\cite{Michael:1986yi,Michael:1995pv}
\begin{equation}\label{MichaelSR1bis}
    M=\frac{\ud\hat\beta}{\ud \ln a}\sum\square,
\end{equation}
where $a$ is the symmetric lattice spacing, $\hat\beta=2N/g^2_0$ is the bare lattice coupling parameter for the $SU(N)$ gauge sector of the theory, and $\sum\square$ is the plaquette action in a one glueball state (with the vacuum value implicitly subtracted) summed over all plaquettes at one time slice. In the naive continuum limit, one can basically write
\begin{equation}
\begin{aligned}
   \frac{\ud\hat\beta}{\ud \ln a}&\to \frac{2\beta(g_0)}{g_0}\,\hat\beta,\\
   \hat \beta\sum\square&\to \langle\Psi|\int\ud^3x\,\frac{1}{4}\,F^2(x)|\Psi\rangle,
\end{aligned}
\end{equation}
so that Eq.~\eqref{MichaelSR1bis} can be interpreted as the lattice version of Eq.~\eqref{traceSR}. We will denote this as
\begin{equation}\label{MichaelSR1}
    M=\frac{\ud\hat\beta}{\ud \ln a}\sum\square\qquad\sim\langle\Psi|\int\ud^3x\,T^\mu_{\phantom{\mu}\mu}(x)|\Psi\rangle.
\end{equation}
As clearly shown by a recent variation of Michael's derivation~\cite{Ji:2021qgo}, the fact that one gets an expression for the mass in terms of the EMT trace follows from an isotropic lattice scaling transformation.

Further relations can be obtained by considering asymmetric lattices. Following the formalism of Ref.~\cite{Karsch:1982ve}, a different lattice coupling parameter $\hat\beta_{\mu\nu}$ must be attributed to the different plaquette orientations $\square_{\mu\nu}$. In the case where one distinguishes the temporal spacing $a_0=a_t$ from the isotropic spatial spacing $a_1=a_2=a_3=a_s$, Michael arrived at the following two sum rules~\cite{Michael:1986yi,Michael:1995pv},
\begin{align}
    M&=\sum (3S\,\square_t+3U\,\square_s)\qquad\sim\langle\Psi|\int\ud^3x\,T^{00}(x)|\Psi\rangle,\label{MichaelSR2}\\
    0&=\sum \left[(2U+S)\,\square_t+(2S+U)\,\square_s\right]\qquad\sim\langle\Psi|\int\ud^3x\,T^{ii}(x)|\Psi\rangle,\label{MichaelSR3}
\end{align}
where $\square_t=\square_{0i}$ and $\square_s=\square_{ij}$ with $i,j\neq 0$. The coefficients $S$ and $U$ are defined at the symmetric point $a_t=a_s=a$ by
\begin{equation}
    \begin{aligned}
       \frac{\partial\hat\beta_{\mu\nu}}{\partial\ln a_\lambda}&=S&&\qquad\text{if}\quad \lambda=\mu\quad\text{or}\quad \lambda=\nu,\\
       \frac{\partial\hat\beta_{\mu\nu}}{\partial\ln a_\lambda}&=U&&\qquad\text{if}\quad \lambda\neq \mu\quad\text{and}\quad \lambda\neq \nu.
    \end{aligned}
\end{equation}
Since Eq.~\eqref{MichaelSR2} is associated with temporal dilatations, it can be interpreted as the lattice version of the energy sum rule~\cite{Ji:2021pys,Ji:2021mtz,Ji:2021qgo}. Similarly, we observe that Eq.~\eqref{MichaelSR3} is associated with spatial dilatations and can therefore be interpreted as the lattice version of the virial theorem. 

Combining Eqs.~\eqref{MichaelSR2} and~\eqref{MichaelSR3}, Michael found two alternative expressions for the glueball mass
\begin{align}
    M&=\sum 2\,(S+U)(3\,\square_t+3\,\square_s)\qquad\sim\langle\Psi|\int\ud^3x\,T^\mu_{\phantom{\mu}\mu}(x)|\Psi\rangle,\label{MichaelSR4}\\
    M&=\sum \frac{2}{3}\,(S-U)(3\,\square_t-3\,\square_s)\qquad\sim\frac{4}{3}\,\langle\Psi|\int\ud^3x\,\bar T^{00}(x)|\Psi\rangle.\label{MichaelSR5}
\end{align}
Using the relation obtained by Karsch~\cite{Karsch:1982ve}
\begin{equation}\label{Karzch1}
    2(S+U)=\frac{\ud\hat\beta}{\ud\ln a}
\end{equation}
and $\square=3\,\square_t+3\,\square_s$, we see that Eq.~\eqref{MichaelSR4} derived from asymmetric lattices and  evaluated at the symmetric point is consistent with Eq.~\eqref{MichaelSR1} obtained directly from symmetric lattices. In the weak-coupling limit $\hat\beta\to\infty$, one finds~\cite{Karsch:1982ve}
\begin{equation}\label{Karsch2}
    2(S-U)\approx-4\hat\beta
\end{equation}
so that one can write~\cite{Michael:1995pv}
\begin{equation}\label{Michael43}
    M\approx\frac{4}{3}\sum\hat\beta(-3\,\square_t+3\,\square_s)\qquad\sim\frac{4}{3}\,\langle\Psi|\int\ud^3x\,\bar T^{00}(x)|\Psi\rangle.
\end{equation}

In the naive continuum limit, the lattice plaquettes are identified with the chromoelectric and chromomagnetic contributions to field energy
\begin{equation}\label{naivelimitEB}
    \begin{aligned}
       \sum\hat\beta(-3\,\square_t)&\to \int\ud^3x\,\frac{\uvec E^2(x)}{2},\\
       \sum\hat\beta(3\,\square_s)&\to \int\ud^3x\,\frac{\uvec B^2(x)}{2}.
    \end{aligned}
\end{equation}
We remind that the sign change in the chromoelectric contribution comes from the transition from Euclidean space to Minkowski space. Using this identification, it appears that the classical form of field energy provides only $3/4$ of the glueball mass~\cite{Michael:1995pv,Rothe:1995av}. Massaging a bit the energy sum rule~\eqref{MichaelSR2}, one has 
\begin{equation}
\begin{aligned}
     M&=\sum \left[\frac{U-S}{2}\,(-3\,\square_t+3\,\square_s)+ \frac{U+S}{2}\,(3\,\square_t+3\,\square_s)\right]\qquad\sim\langle\Psi|\int\ud^3x\,T^{00}(x)|\Psi\rangle\\
     &\approx\sum \left[\hat\beta(-3\,\square_t+3\,\square_s)+\frac{1}{4}\,\frac{\ud\hat\beta}{\ud\ln a}\,(3\,\square_t+3\,\square_s)\right].
     \end{aligned}
\end{equation}
In the naive continuum limit, Rothe concluded that the missing $1/4$ of the glueball mass comes from the trace anomaly, in apparent agreement with the analysis of Refs.~\cite{Ji:1994av,Ji:1995sv}. The recent works~\cite{Ji:2021pys,Ji:2021mtz,Ji:2021qgo} present a variation of this discussion and use it as a support of the concept of QAE.

This result, however, has to be considered with a grain of salt, since it relies on both the weak-coupling limit~\eqref{Karsch2} and the naive continuum limit~\eqref{naivelimitEB}. Non-perturbative evaluations at finite temperature of the combination $2(S-U)$ show in fact significant deviations from the weak-coupling value~\cite{Engels:1994xj,Engels:1999tk,Meyer:2007ic}. This suggests that in general the classical form of field energy on the lattice does not provide exactly $3/4$ of the glueball mass, and indicates that the lattice operators must be renormalized. This is to be expected since $(-3\,\square_t+3\,\square_s)$ is not a Noether charge due to the breaking of Poincar\'e symmetry on the lattice, where the hypercubic symmetry typically leads to more complicated mixing patterns than in the continuum~\cite{Caracciolo:1991cp,Capitani:1994qn,Gockeler:1996mu,Meyer:2007fc,Meyer:2007ic}. In particular, the chromoelectric and chromomagnetic contributions to the field energy mix under renormalization~\cite{Bali:1994de,Gockeler:1996gu}, invalidating therefore the naive continuum limit interpretation~\eqref{naivelimitEB}.

In conclusion, contrary to the suggestion of Refs.~\cite{Ji:2021pys,Ji:2021mtz,Ji:2021qgo} a careful inspection reveals that the lattice energy sum rule does not provide a clear support to the concept of QAE. In particular, it is essential to consider the renormalization of the lattice operators before providing any physical interpretation. (Note that $\uvec E^2$ and $\uvec B^2$ in Eq.~\eqref{naivelimitEB} do not have the same meaning as the corresponding renormalized operators in the continuum.) This is crucial for the components of the EMT since the breaking of Poincar\'e symmetry on the lattice is a source of artifacts, as we will see in the following.

\subsection{Translation symmetry}

How to construct the EMT on the lattice is a tough question that has been studied for over 30 years~\cite{Caracciolo:1988hc,Caracciolo:1989bu,Caracciolo:1989pt,Caracciolo:1991vc,Caracciolo:1991cp,Buonanno:1995us,Caracciolo:2000td,Suzuki:2013gza,DelDebbio:2013zaa,Makino:2014taa,Giusti:2015daa,Harlander:2018zpi,Yanagihara:2018qqg,Yanagihara:2019foh,Yanagihara:2020tvs,DallaBrida:2020gux,Panagopoulos:2020qcn,Suzuki:2021tlr}. For recent investigations of the hadron mass structure on the lattice see also Refs.~\cite{Yang:2014xsa,Yang:2018nqn,Sun:2020pda,He:2021bof,Liu:2021gco}. A major difficulty is that lattice regularization breaks translation symmetry and makes the construction and renormalization of the EMT non-trivial. In particular, it turns out that any discretization of the classical EMT, denoted $T^{\mu\nu}_\text{tree}$, is not conserved in the quantum theory~\cite{Caracciolo:1988hc,Caracciolo:1989bu,Caracciolo:1989pt,Caracciolo:1991vc,Caracciolo:1991cp},
\begin{equation}
    \partial_\mu T^{\mu\nu}_\text{tree}=R^\nu+X^\nu.
\end{equation}
Here $R^\nu$ is proportional to the lattice EOM and $X^\nu$ is an operator that formally vanishes when the lattice spacing $a$ tends to zero. Because of radiative corrections, $X^\nu$ provides however finite contributions when inserted into Green functions. It vanishes for zero external momentum transfer and can thus be rewritten as $X^\nu=-\partial_\mu T^{\mu\nu}_\text{corr}$. The conserved EMT on the lattice is therefore defined as
\begin{equation}\label{correctedH}
    T^{\mu\nu}=T^{\mu\nu}_\text{tree}+T^{\mu\nu}_\text{corr}.
\end{equation}
The correction term $T^{\mu\nu}_\text{corr}$ ensures that the translational Ward identities are satisfied. At the same time it ensures that the trace anomaly is correctly reproduced\footnote{Interestingly, a similar structure was found in Ref.~\cite{Hata:1980fq}  using Fujikawa's method to derive the trace anomaly from a path-integral approach.}. The fact that the classical expression for the EMT must be corrected is consistent with the general observation that once a genuine symmetry is broken by the regulator, one should expect to see the appearance of additional symmetry-restoring counterterms~\cite{Bogolyubov:1959bfo,tHooft:1971akt,Hata:1980fq,Belusca-Maito:2020ala}.

In the recent work~\cite{Ji:2021qgo}, the mass structure of a $(1+1)$-dimensional non-linear sigma model in the large-$N$ limit is studied at the one-loop level. It is found that the operator form of the total Hamiltonian $H=\int\ud^3x\,T^{00}(x)$ depends on the choice of regularization scheme. In particular, the classical Hamiltonian $H_c=\int\ud^3x\,T^{00}_\text{tree}(x)$ is regulator-dependent and has no universal physical meaning. The authors observe that in symmetric regularization schemes, where all directions are treated equally, the classical Hamiltonian coincides with the traceless part $H_T=\int\ud^3x\,\bar T^{00}(x)$, while in regularization schemes where the energy integral can be rescaled back and forth (like e.g.~dimensional regularization) the classical Hamiltonian coincides with the total Hamiltonian. 

These results can easily be understood from the point of view of translation symmetry. Regularization schemes where the energy integral can be rescaled back and forth are precisely those preserving translation symmetry in the temporal direction. It should therefore not be surprising that the total Hamiltonian takes the same form as in the classical theory, since this is a mere consequence of the quantum action principle~\cite{Breitenlohner:1977hr,Collins:1984xc}. More generally, it has been argued that translational Ward identities obtained with a cutoff procedure preserving Poincar\'e symmetry cannot contain anomalies~\cite{Coleman:1970je,Zacrep:1975ds}. On the other hand, when translation invariance in the temporal direction is broken by the regulator, the total Hamiltonian must necessarily involve additional contributions like in Eq.~\eqref{correctedH} to restore translation symmetry. These correction terms should be regarded as mere artifacts arising due to a poor choice of symmetry-breaking regulator, and not as genuine physical contributions\footnote{Similarly, if the regulator breaks gauge symmetry one usually needs to include a gauge boson mass counterterm to restore the symmetry~\cite{tHooft:1971akt}, and in particular to ensure that the physical gauge boson mass vanishes.}. They must disappear in the process of renormalization to comply with Poincar\'e symmetry and the quantum action principle.

The authors of Ref.~\cite{Ji:2021qgo} observed that the one-loop contribution to the classical Hamiltonian vanishes only in symmetric regularization schemes. In other schemes, the naively vanishing integral gives a finite contribution because of the asymmetric regulator, which has been interpreted as a sign of their anomalous nature. In particular, in dimensional regularization the one-loop integral takes the form
\begin{equation}\label{DRmech}
    \int\frac{\ud k_4\,\ud^{1-2\epsilon}k_1}{(2\pi)^{2-2\epsilon}}\,\frac{k^2_4-k^2_1}{(k^2+m^2)^2}=\frac{\epsilon}{1-\epsilon}\int\frac{\ud^{2-2\epsilon}k}{(2\pi)^{2-2\epsilon}}\,\frac{k^2}{(k^2+m^2)^2}.
\end{equation}
The symmetric integral generates a $\frac{1}{4\pi\epsilon}$ pole that cancels the explicit $\epsilon$ factor in front of it, leading to a finite result. Since a similar mechanism is responsible for the trace anomaly in dimensional regularization, the authors of Ref.~\cite{Ji:2021qgo} interpreted the non-vanishing of the one-loop contribution to the classical Hamiltonian as of anomalous nature.

We disagree with this interpretation. The trace anomaly arises in dimensional regularization from an evanescent term of the form $\epsilon O$ with $O$ some operator. At the classical level, the operator $O$ is finite so that the evanescent term vanishes in the limit $\epsilon\to 0$. At the quantum level, the operator contains a $\frac{1}{\epsilon}$ pole leading to the trace anomaly. In contradistinction, $H_c$ is not an evanescent operator and does not vanish at the classical level in the limit $\epsilon\to 0$. The explicit $\epsilon$ factor is not part of the definition of $H_c$ but appears in Eq.~\eqref{DRmech} only after a change of variables. Despite the similitude with the trace anomaly mechanism, the one-loop contribution to $H_c$ is actually not anomalous. 

Anomalies are usually associated with a pair of symmetries~\cite{Morozov:1986hv}. In the present case, the pair consists of translation and dilatation symmetries. In the regularized theory, there is no way of preserving at the same time both translation symmetry and the standard definition of trace. In dimensional regularization, Poincar\'e symmetry is preserved by distorting the classical $d=4$ spacetime into a $d=4-2\epsilon$ one, affecting therefore the definition of trace; see also Appendix~\ref{App:DR}. Pauli-Villars regularization also preserves Poincar\'e symmetry but adds regulator fields which provide new contributions to the trace. Lattice regularization, on the other hand, preserves the classical $d=4$ spacetime but breaks Poincar\'e symmetry by making it discrete. In this case the definition of trace is unaffected by the regularization and the trace anomaly must appear in the form of correction terms to the EMT ensuring that the translational Ward identities are satisfied. 

To sum up, the appearance of trace anomaly contributions to the energy on the lattice is a pure artifact associated with the breaking of translation symmetry by the discretization.  Translation symmetry being an exact symmetry at the quantum level, this artifact should disappear in the process of renormalization to agree with the results obtained using symmetry-preserving regularization schemes. In conclusion, we do not find any concrete support to the concept of QAE from LQCD.

\section{Conclusions}\label{sec:conclusions}
Understanding the decomposition of the nucleon mass in QCD in terms of contributions from quarks and gluons is a topic of high interest and fundamental importance.  
Presently, different opinions exist in this area, reflected by different mass decompositions in the literature. 
Here we have concentrated on the mass decompositions which are based on the component $T^{00}$ of the QCD EMT: a four-term decomposition originally proposed in Refs.~\cite{Ji:1994av,Ji:1995sv} and recently slightly modified in Refs.~\cite{Bali:2016lvx,Yang:2018nqn,Liu:2021gco, Ji:2021pys, Ji:2021mtz,Ji:2021qgo}, a two-term decomposition put forward in Ref.~\cite{Lorce:2017xzd}, and a three-term decomposition arrived at in Refs.~\cite{Rodini:2020pis,Metz:2020vxd}.  
The latter two are very closely related --- by separating the total quark contribution to the nucleon mass into quark kinetic plus potential energies and a quark mass term one obtains the three-term decomposition starting from the two-term decomposition.

One controversy concerns the proper expressions for the renormalized operators of the mass decomposition. We have elaborated on this important point using DR and MS-type schemes, and we re-confirm the findings of Refs.~\cite{Rodini:2020pis,Metz:2020vxd} in that regard, which to some extent are based on the renormalization of the full EMT discussed in Refs.~\cite{Hatta:2018sqd,Tanaka:2018nae}. 
This implies, in particular, that in DR the operator $\frac{1}{2}(\uvec E^2 +\uvec B^2)_R$ has a unique meaning in terms of components of the EMT.
This operator corresponds to the total gluon contribution to the nucleon mass.
Furthermore, different points of view exist with regard to the physical interpretation of the terms in the mass decompositions.
We re-iterate the concern raised in Ref.~\cite{Lorce:2017xzd} that the four-term decomposition contains mixtures of genuine energy terms and pressure-volume terms.
This feature is closely related to the fact that, in order to derive the four-term decomposition, one must make use of the condition for mechanical equilibrium of the  nucleon.
As we have shown, this condition actually coincides with the virial theorem, which we have discussed at length.
Both the two-term and the three-term decomposition do not make use of the virial theorem, and their contributions have a clean physical interpretation.
One argument that was put forth in favor of the four-term decomposition is that it contains the so-called ``quantum anomalous energy'', which has been suggested as a unique contribution to the nucleon mass~\cite{Ji:2021pys,Ji:2021mtz,Ji:2021qgo}.
We have explained why, in our view, this term is not a genuine contribution to the mass decomposition.

Even though the two-term and three-term mass decompositions do not contain the operator of the trace anomaly, it remains important to pursue attempts to measure (the gluon contribution to) the trace anomaly~\cite{Kharzeev:1995ij,Meziani:2020oks,Hatta:2018ina,GlueX:2019mkq,Hatta:2019lxo,Mamo:2019mka,Wang:2019mza,Boussarie:2020vmu,Gryniuk:2020mlh,Wang:2021ujy,Zeng:2020coc,Wang:2021dis,Sun:2021gmi,Xie:2021seh}. 
Such measurements can help obtaining a more robust phenomenology of the quark and gluon contributions to the EMT trace. 
This in turn allows one to better pin down the quark mass term and as such the numerics of all the terms of the nucleon mass decomposition. 
Finally, we would like to emphasize that 
all the nucleon mass decompositions require the same phenomenological input, namely  
two independent gravitational form factors.

\begin{acknowledgements}
We would like to dedicate this work to the memory of Maxim Polyakov, who sadly passed away during the preparation of this manuscript. C.~L.~is grateful to Maxim Polyakov, Oleg Teryaev, and Christoph Kopper for illuminating discussions. The work of A.~M.~has been supported by the National Science Foundation under grant number PHY-2110472 and by the U.S. Department of Energy, Office of Science, Office of Nuclear Physics, within the
framework of the TMD Topical Collaboration. 
The work of B.~P. and S.~R. is part of a project that has received funding from the European Union’s Horizon 2020 research and innovation programme under grant agreement
STRONG - 2020 - No 824093.

\end{acknowledgements}

\appendices

\section{Virial theorem}\label{App:virial}

In this Appendix, we review the virial theorem in various contexts and discuss its physical meaning.

\subsection{Classical point mechanics}

Originally, the virial theorem comes from classical point mechanics~\cite{Clausius:1870} where one considers a system of discrete pointlike particles bound by potential forces. Denoting by $\uvec r_k$ and $\uvec p_k$ the position and momentum of the $k$th particle, one introduces the quantity
\begin{equation}
    G=\sum_k \uvec p_k\cdot\uvec r_k
\end{equation}
whose time derivative can be expressed as
\begin{equation}\label{dGdTmech}
    \frac{\ud G}{\ud t}=\sum_k\left(\uvec p_k\cdot\uvec v_k+\uvec F_k\cdot\uvec r_k\right),
\end{equation}
where the velocity of the $k$th particle is defined as $\uvec v_k=\ud\uvec r_k/\ud t$ and the net force acting on it as $\uvec F_k=\ud\uvec p_k/\ud t$. In both relativistic and non-relativistic cases, the velocity of a particle can be expressed as the derivative of the kinetic energy with respect to momentum. Moreover, if the forces derive from a potential that depends only on the coordinates, we can finally write
\begin{equation}\label{virialgeneral}
    \frac{\ud G}{\ud t}=\sum_k\left(\uvec p_k\cdot\tfrac{\partial}{\partial\uvec p_k}\ucal T(\{\uvec p_i\})-\uvec r_k\cdot\tfrac{\partial}{\partial\uvec r_k}\ucal V(\{\uvec r_i\})\right),
\end{equation}
where $\ucal T(\{\uvec p_i\})$ and $\ucal V(\{\uvec r_i\})$ are the total kinetic and potential energies depending on all the momentum and position variables, respectively. 

For convenience, we introduce a double square bracket notation
\begin{equation}
    [\![O]\!]\equiv\lim_{\tau\to\infty}\frac{1}{\tau}\int_0^\tau\ud t\,O(t)
\end{equation}
to indicate that some quantity $O$ is averaged over a long time. One can then write
\begin{equation}
    \bigg[\!\!\bigg[\frac{\ud G}{\ud t}\bigg]\!\!\bigg]=\lim_{\tau\to\infty}\frac{G(\tau)-G(0)}{\tau}.
\end{equation}
Now, for a bound system in the center-of-mass frame\footnote{It seems that the condition on the reference frame is often omitted in the literature. It is however essential since $G(t)$ receives a contribution from the center-of-mass motion that grows linearly with $t$ for a closed system, unless the total momentum vanishes.}, particle coordinates and momenta are expected to be bounded, so that $G_\text{min}\leq G(t)\leq G_\text{max}$ for all $t$ with both $G_\text{min}$ and $G_\text{max}$ finite. In that case, one expects $[\![\ud G/\ud t]\!]=0$ and hence
\begin{equation}\label{virialphasespace}
    \sum_k\Big[\!\!\Big[\uvec p_k\cdot\tfrac{\partial}{\partial\uvec p_k}\ucal T\Big]\!\!\Big]= \sum_k\Big[\!\!\Big[\uvec r_k\cdot\tfrac{\partial}{\partial\uvec r_k}\ucal V\Big]\!\!\Big].
\end{equation}
This is a generic form of the virial theorem in classical point mechanics, valid for both relativistic and non-relativistic theories~\cite{Lucha:1989jf,Lucha:1990mm}.

In particular, for a non-relativistic theory with a potential between any two particles $i$ and $j$ of the form $\ucal V(\uvec r_i,\uvec r_j)=Cr^n$, where $r$ is the relative distance and $C$ is some constant, the virial theorem reduces to a simple relation between the time-averaged total kinetic and potential internal energies
\begin{equation} \label{e:virial_cl_T_V}
    2[\![\ucal T]\!]=n[\![\ucal V]\!],
\end{equation}
and allows one to express e.g. the total center-of-mass energy purely in terms of $[\![\ucal V]\!]$. The kinetic energy being always positive, the sign of the constant $C$ must be the same as the sign of $n$ for the bound system to exist. In other words, the net forces must be attractive.

More generally, one can write the virial theorem in non-relativistic point mechanics as
\begin{equation}
    [\![\ucal T]\!]=-\frac{1}{2}\sum_k[\![ \uvec r_k\cdot\uvec F_k]\!].
\end{equation}
The quantity on the right-hand side is called the \emph{virial}, derived from the latin word \emph{vis} meaning ``force'', ``energy'' or ``power''. This form of the virial theorem is very useful e.g. for the description of gases. Indeed, for a non-relativistic gas contained in a box of volume $V$ at rest with a constant pressure $p$, the virial theorem tells us that~\cite{Schectman:1957}
\begin{equation} \label{virial_general}
    [\![\ucal T]\!]=\frac{3}{2}\,pV-\frac{1}{2}\sum_k[\![ \uvec r_k\cdot\uvec F^\text{int}_k]\!],
\end{equation}
where the second term on the r.h.s.~is the virial associated with internal forces only, and the first term corresponds to the contribution arising from the external forces exerted by the walls of the box 
\begin{equation}
-\frac{1}{2}\sum_k[\![ \uvec r_k\cdot\uvec F^\text{ext}_k]\!]=\frac{1}{2}\,p\int\ud\uvec S\cdot\uvec r=\frac{1}{2}\,p\int\ud^3r\,\uvec\nabla\cdot\uvec r=\frac{3}{2}\,p V.    
\end{equation}
For ideal gases there are no internal forces and so $ [\![\ucal T]\!]=\frac{3}{2}\,pV$. If the gas is composed of $N$ particles, the equipartition theorem tells us that $[\![\ucal T]\!]=\frac{3}{2}\,Nk_BT$, where $k_B$ is the Boltzmann constant and $T$ is the temperature, and one obtains finally the ideal gas law $pV=Nk_BT$.

\subsection{Quantum mechanics}\label{App:virialMQ}

The virial theorem also holds in quantum mechanics~\cite{Born:1926uzf,Finkelstein:1928,Fock:1930}. Indeed, similarly to classical mechanics one defines the operator $G$ as\footnote{In the literature, one uses sometimes a non-symmetric definition $G=\sum_k\uvec p_k\cdot\uvec r_k$ in quantum mechanics. The ambiguity associated with the ordering of operators will however be irrelevant. Indeed, the various orderings differ only by a term proportional to the identity operator and give therefore the same commutators of $G$ with some other operator, which are the only quantities needed for our presentation.}
\begin{equation}
    G=\frac{1}{2}\sum_k(\uvec r_k\cdot\uvec p_k+\uvec p_k\cdot\uvec r_k).
\end{equation}
According to Erhenfest's theorem, we can write
\begin{equation}
    \frac{\ud}{\ud t}\langle G\rangle=\frac{1}{i}\langle[G,H]\rangle
\end{equation}
with $\langle O\rangle$ the expectation value associated with some normalizable state. For a Hamiltonian of the form
\begin{equation}
    H=\sum_k\frac{\uvec p^2_k}{2m_k}+\ucal V(\{\uvec r_k\})
\end{equation}
we get
\begin{equation}
    \frac{\ud}{\ud t}\langle G\rangle=2\langle\ucal T\rangle-\sum_k\langle\uvec r_k\cdot\tfrac{\partial}{\partial\uvec r_k}\ucal V\rangle
\end{equation}
with $\ucal T=\sum_k\,\uvec p^2_k/2m_k$ the usual non-relativistic kinetic energy operator. 

Now, for a normalizable stationary state $H|\Psi\rangle=E|\Psi\rangle$ we have $\ud\langle\Psi| G|\Psi\rangle/\ud t=0$, leading to the quantum version of the non-relativistic virial theorem
\begin{equation}\label{virialQ}
 2\langle\Psi|\ucal T|\Psi\rangle=\sum_k\langle\Psi|\uvec r_k\cdot\tfrac{\partial}{\partial\uvec r_k}\ucal V|\Psi\rangle. 
\end{equation}
If the potential between two particles is of the form $\ucal V(\uvec r_i,\uvec r_j)=Cr^n$, Eq.~\eqref{virialQ} becomes $2 \langle\Psi| \ucal T|\Psi \rangle = n \langle \Psi|\ucal V|\Psi \rangle$ which is the quantum-mechanical counterpart of Eq.~\eqref{e:virial_cl_T_V}.
Denoting the center-of-mass position and momentum operators by $\uvec r_\text{CM}$ and $\uvec p_\text{CM}$, we have also $\uvec 0=\ud\langle\Psi| \uvec r_\text{CM}|\Psi\rangle/\ud t=\langle\Psi|\uvec p_\text{CM}|\Psi\rangle/m$ indicating that a stationary state is in average at rest\footnote{One may argue that momentum eigenstates with non-vanishing momentum are also eigenstates of the Hamiltonian. However, these states are non-normalizable and one cannot automatically conclude that $\ud\langle G\rangle/\ud t=0$.}. This is consistent with the observation that Eq.~\eqref{virialQ} cannot be valid in all frames, since for a closed bound system the total kinetic energy increases with the total momentum of the system, whereas the potential energy does not~\cite{Killingbeck:1970}. Note also that if we work with a superposition of stationary states, the expectation value $\langle\Psi| G|\Psi\rangle$ will generally depend on time. However, like in the classical case we may still expect it to be bounded, so that $[\![\ud\langle\Psi| G|\Psi\rangle/\ud t]\!]=0$~\cite{Crawford:1989}.

It was later realized that $G$ is simply the generator of spatial dilatations $U_D=e^{-i\kappa G}$, as one can see from
\begin{equation}
    U^{-1}_D\uvec r_kU_D=\lambda\uvec r_k,\qquad U^{-1}_D\uvec p_kU_D=\lambda^{-1}\uvec p_k
\end{equation}
with $\lambda=e^\kappa$. This is also clear from
\begin{equation}\label{HGcomm}
    \frac{1}{i}[H,G]=-\sum_k\uvec p_k\cdot\tfrac{\partial}{\partial\uvec p_k}\ucal T+\sum_k\uvec r_k\cdot\tfrac{\partial}{\partial\uvec r_k}\ucal V=\sum_k\left(\uvec r_k\cdot\tfrac{\partial}{\partial\uvec r_k}-\uvec p_k\cdot\tfrac{\partial}{\partial\uvec p_k}\right)H,
\end{equation}
since $\uvec p_k\cdot\tfrac{\partial}{\partial\uvec p_k}$ and $\uvec r_k\cdot\tfrac{\partial}{\partial\uvec r_k}$ measure the degree of homogeneity in momentum and position space, respectively. This observation leads to an interesting alternative derivation of the virial theorem from a variational approach~\cite{Fock:1930,Puff:1968,Killingbeck:1970,Gersch:1979,Kleban:1979} which applies also to relativistic quantum mechanics. Let us introduce the function
\begin{equation}
    E(\kappa)=\langle\Psi_\kappa|H|\Psi_\kappa\rangle=\langle\Psi|U^{-1}_DHU_D|\Psi\rangle,
\end{equation}
where $|\Psi\rangle$ is some state and $|\Psi_\kappa\rangle=U_D|\Psi\rangle$ is its dilated counterpart which depends on the parameter $\kappa$. In a variational approach, we require that the eigenvalue $E$ of a stationary state $|\Psi\rangle$ must be an extremum of $E(\kappa)$ at $\kappa=0$. In other words, we demand that
\begin{equation}
    \begin{aligned}
    E(0)&=E,\\
    \frac{\partial E}{\partial\kappa}(0)&=\langle\Psi|\frac{\partial (U^{-1}_DHU_D)}{\partial\kappa}\bigg|_{\kappa=0}|\Psi\rangle=0.
    \end{aligned}
\end{equation}
Using Eq.~\eqref{HGcomm}, the stationarity condition under spatial rescaling gives directly the virial theorem
\begin{equation}\label{virialQM}
   \sum_k\langle\Psi| \uvec p_k\cdot\tfrac{\partial }{\partial\uvec p_k}\ucal T|\Psi\rangle=\sum_k\langle\Psi|\uvec r_k\cdot\tfrac{\partial }{\partial\uvec r_k}\ucal V|\Psi\rangle,
\end{equation}
recognized as the quantum-mechanical counterpart of Eq.~\eqref{virialphasespace}. 

The connection between this derivation and the former one simply follows from the identity
\begin{equation}\label{relid}
    \frac{\ud G}{\ud t}=\frac{1}{i}[G,H]=-\frac{1}{i}[H,G]=-\frac{\partial( U^{-1}_DHU_D)}{\partial\kappa}\bigg|_{\kappa=0}
\end{equation}
which relates the breaking of dilatation symmetry to the behavior of the Hamiltonian under dilatations. This shows that the virial theorem is fundamentally a statement about mechanical equilibrium expressed by the stationarity of the system under spatial dilatations. Note however that it does not say anything about stability since the latter is determined by the second derivative w.r.t.~$\kappa$.

The virial theorem is often used to simplify the calculation of the total energy of a bound system.  Let us consider for example a relativistic spin-$1/2$ particle in a static external potential~\cite{Rafelski:1977vq}. In the Dirac theory, we can write the particle energy as
\begin{equation}
    E=\int\ud^3r\,\psi^\dag\uvec r\cdot\uvec\nabla V(\uvec r)\psi+m\int\ud^3r\,\psi^\dag\beta \psi+\int\ud^3r\,\psi^\dag V(\uvec r)\psi,
\end{equation}
since the virial theorem~\eqref{virialQM} tells us that $\int\ud^3r\,\psi^\dag\uvec\alpha\cdot\uvec p\,\psi =\int\ud^3r\,\psi^\dag\uvec r\cdot\uvec\nabla V(\uvec r)\psi$. For a Coulomb potential $V_C(r)\propto 1/r$, we have $\uvec r\cdot\uvec\nabla V_C(r)=-V_C(r)$ so that we get the remarkably simple expression
\begin{equation} \label{virial_simple}
    E=m\int\ud^3r\,\psi^\dag\beta \psi.
\end{equation}
Evaluating the r.h.s.~of Eq.~\eqref{virial_simple} with a trial wave function provides, in a simple manner, an upper bound for the energy of the system.
For a free particle, the rest energy is $E=m$ and we recover from Eq.~\eqref{virial_simple} the expected normalization of the free Dirac wave function $\int\ud^3r\,\psi^\dag\beta \psi=1$. For a bound particle, we expect $E<m$ and hence $\int\ud^3r\,\psi^\dag\beta \psi<1$.

\subsection{Link between field theory and point mechanics}\label{App:fieldtheory}

As discussed in Sec.~\ref{sec:virialthmfield} the virial theorem can be extended to a field-theoretical framework. We show here how the continuum treatment reduces to the usual one in point mechanics. For a point particle we can write the momentum density as $T^{0i}(x)=p^i(x)\,\delta^{(3)}(\uvec x-\uvec r(t))$. Its time derivative receives therefore two contributions,
\begin{equation}\label{Eulerdt}
    \partial_0T^{0i}(x)=\partial_0p^i(x)\,\delta^{(3)}(\uvec x-\uvec r(t))-p^i(x)\,\uvec v(t)\cdot\uvec\nabla_{\!x}\delta^{(3)}(\uvec x-\uvec r(t)),
\end{equation}
where $\uvec v=\ud\uvec r/\ud t$ is the particle velocity. The net force acting on the particle is obtained by integrating this equation over space
\begin{equation}
    F^i(t,\uvec r(t))=\frac{\ud}{\ud t}\int\ud^3x\,T^{0i}(x)=\frac{\ud }{\ud t}p^i(t,\uvec r(t))=\frac{\partial}{\partial t}p^i(t,\uvec r(t))+\uvec v(t)\cdot\uvec\nabla p^i(t,\uvec r(t)).
\end{equation}
We recognize as expected the expression for the material derivative. We can then rewrite Eq.~\eqref{Eulerdt} as
\begin{equation}\label{Eulerbis}
     \partial_0T^{0i}(x)=F^i(t,\uvec r(t))\,\delta^{(3)}(\uvec x-\uvec r(t))-\uvec v(t)\cdot\uvec\nabla T^{0i}(x).
\end{equation}
So the rate of momentum change $\partial_0T^{0i}(x)$ at a fixed spacetime location (Eulerian description) is related to the rate of momentum change $F^i(t,\uvec r(t))\,\delta^{(3)}(\uvec x-\uvec r(t))=\frac{\ud}{\ud t}T^{0i}(x)$ of the fixed material point (Lagrangian description) via the convective rate of momentum change $\uvec v(t)\cdot\uvec\nabla T^{0i}(x)$. Remembering now that $G=\sum_i\int\ud^3x\,T^{0i}x^i$, we get
\begin{equation}
    G(t,\uvec r(t))=\uvec r(t)\cdot\uvec p(t,\uvec r(t))
\end{equation}
and
\begin{equation}
    \frac{\ud }{\ud t} G(t,\uvec r(t))=\uvec r(t)\cdot\uvec F(t,\uvec r(t))+\uvec v(t)\cdot\uvec p(t,\uvec r(t))
\end{equation}
which agree with the expressions found in point mechanics, where the explicit and implicit time dependences of $G$, $\uvec F$ and $\uvec p$ are merged into a single total time dependence. 

Comparing now Eq.~\eqref{Eulerbis} with Eq.~\eqref{id1} for $\mu=i$, we find that
\begin{equation}
    \mathcal F^i(x)=F^i(t,\uvec r(t))\,\delta^{(3)}(\uvec x-\uvec r(t)),
\end{equation}
as expected, but also
\begin{equation}
    T^{ki}(x)=v^k(t)T^{0i}(x)
\end{equation}
which indicates that the stress tensor associated with a point particle is simply given by the tensor product of the momentum density with the velocity. Put differently, the stress tensor arises from the sole motion of the particle and has therefore purely convective contributions.
Since a point particle has no extension, it has no internal, that is, rest-frame pressure.

\section{Dimensional regularization}\label{App:DR}

In this section, we summarize the main working principles of dimensional regularization (DR), as described in  the  works of Refs. [82,84]. In particular, we will outline the properties of the (infinite-dimensional) domain space in the
DR approach and how  the physical space of a system can consistently be incorporated as a subspace of this domain.

DR is defined only in perturbation theory through the following general procedure: for any given Green's function, one inserts an  expansion (of order $n$) of the exponential of the action and  replaces the usual momentum integration in $4$ dimensions with a map $\mathcal{I}_d$ which is usually called `integration in $d$ dimensions'. Since the momentum integration has changed, it is natural that also the `momenta' have changed their nature, from four-vectors to objects with different dimensionality. These objects actually take different definitions depending on the type of DR that one adopts. 
We can distinguish the following DR types (see, e.g., Ref.~\cite{Gnendiger:2017pys}, for a comprehensive overview):
the conventional DR (CDR), where one treats all the vectors and tensors in $d$ dimensions (see below for the proper definition of $d$-dimensional vectors); the 't Hooft-Veltman (HV) regularization, where one attributes  $d$ dimensions only to the `singular' or `internal' vector (or tensor) fields and $4$ dimensions to all other fields;  the four-dimensional helicity scheme (FDH) and dimensional reduction (DRED), where one treats the momenta in $d$ dimensions and  enlarges the space to  $d_s=d+n_\epsilon$ dimensions to treat the singular vector fields in $d_s$ dimension, working with the regular vector fields in $4$ (within FDH) or $d_s$ (within DRED) dimensions.

A common misconception is that any DR scheme extends or contracts the Minkowski momentum space S$_4$ into a non-integer vector space. This is not how DR actually works at a consistent theoretical level. 
DR extends S$_4$ into an \textit{infinitely-dimensional} vector space, indicated as a quasi-$d$-dimensional space\footnote{Usually one sets $d=4-2\epsilon$} $\text{QS}_d$, which may be further enlarged to a quasi-$d_s$ dimensional space, denoted by $\text{QS}_{d_s}$, via a direct (orthogonal) sum with $\text{QS}_{n_\epsilon}$,
\begin{eqnarray}
\text{QS}_{d_s}=\text{QS}_{d}\oplus\text{QS}_{n_\epsilon}, \quad \text{S}_4\subset \text{QS}_d.
\end{eqnarray}
The space $\text{QS}_{d}$ is the natural domain of CDR and of momentum
integration in all the DR schemes, and the claim that one works with $d$ dimensions comes from the scaling property of the map $\mathcal{I}_d$, which resembles the scaling property of a finite dimensional space with $d$ dimensions.
We can briefly summarize the properties of the map $\mathcal{I}_d$ 
which is at the 
core of all the DR variants following the outline in Refs.~\cite{Collins:1984xc,Wilson:1972cf}.
Formally, $\mathcal{I}_d$ is defined for scalar `integrands' as the map from the direct product of space functions on $\text{QS}_d$ and $\text{QS}_d$ into complex numbers, i.e.,
\begin{align}
\mathcal{I}_d: \, \mathcal{F}(\text{QS}_d)\otimes \text{QS}_d &\rightarrow \mathbb{C}, \notag \\
 (f,\bm{p}) &\mapsto \mathcal{I}_d (f,\bm{p}).
\end{align}
In the following, we will assume that $\text{QS}_d$ is a Euclidean space. To translate the theory from Minkowski space into Euclidean space one can either apply a Wick-rotation  from the very beginning or can single out the time dimension by writing $\bm{p} = (p^0,\bm{q})$
\[
\Id(f,\bm{p}) = \int dp^0 \ \mathcal{I}_{d-1}(f(p^0,\bm{q}),\bm{q}).
\]
The separation of $\Id$ into a standard one-dimensional integral and a residual $d-1$ map is allowed and consistent, as it will become evident from the discussion below. This is what is meant when working in 1+(3-2$\epsilon$) dimensions~\footnote{Similarly, using light-front variables, the choice $d= 2+(2-2\epsilon) $ implies an integration over  the $\pm$ components in Eq. (171).}.

The fundamental axioms for the map $\Id$ are:
\begin{enumerate}
    \item Linearity: $\Id(af+bg,\bm{p}) = a\Id(f,\bm{p})+b\Id(g,\bm{p});$
    \item Scaling: $\Id(f,s\bm{p}) = s^{-d}\Id(f,\bm{p}),$ where this property sets the `dimension' of the spacetime to $d$;
    \item Translational invariance: $\Id(f(\bm{p}+\bm{q}),\bm{p}) = \Id(f,\bm{p});$
    \item Rotational invariance: $\Id(f(R\bm{p}),\bm{p}) = \Id(f,\bm{p}),$  with $R$ being an arbitrary rotation matrix in $\text{QS}_d$.
\end{enumerate}
From these axioms, one has to prove the existence and uniqueness of the map $\Id$.
We start with the uniqueness.
Suppose that the map $\Id$ exists. From the linearity axiom, we can expand any function in $\mathcal{F}(\text{QS}_d)$ in terms of a set of basis functions which can be taken as~\cite{Wilson:1972cf}
\[
f_{s,\bm{q}}(\bm{p}) = f(s(\bm{p}+\bm{q})) =\exp\ta -s^2(\bm{p}+\bm{q})^2\tc.
\]
Hence, $\Id(g,\bm{p})$ can be expressed, for any function $g$, as a superposition of the master integral
\[
\Id(e^{-\bm{p}^2},\bm{p}) ,
\]
which is set \textit{by definition} to
\[
\Id(e^{-\bm{p}^2},\bm{p}) = \pi^{d/2}.
\]
Therefore, up to a redefinition of the normalization of $\Id$, if the map exists, it is also unique.

To prove the existence, one first assumes that all the external physical vectors $\bm{q}_i$   in $\text{QS}_d$  belong to a vector subspace $V\subset \text{QS}_d$ with dim$(V)=J<+\infty$. This assumption is not restrictive, since any external vector should live in the physical Minkowski space of the theory (for standard QED or QCD the physical space is $4$-dimensional). Any vector $\bm{p}\in \text{QS}_d$ can then  be written as  $\bm{p} = \bm{p}_\para + \bm{p}_\perp$, where $\bm{p}_\para\in V$ and $\bm{p}_\para\cdot \bm{p}_\perp = 0$.
Within such a decomposition, a generic scalar function is  given by
\[
f(\bm{p},\{\bm{q}_i\}_i) \equiv f(p^2, \{\bm{p}\cdot \bm{q}_i\}_i,\{\bm{q}_i\cdot\bm{q}_j\}_{i,j}),
\]
and 
one \textit{defines} the map $\Id$ to be the ordinary $J$-dimensional integral over $\bm{p}_\para$ performed after the integration in one dimension over $p_\perp = |\bm{p}_\perp|$ with a weight $p_\perp^{d-J-1}$, i.e.,
\[
\Id\ta f(p^2, \{\bm{p}\cdot \bm{q}_i\}_i,\{\bm{q}_i\cdot\bm{q}_j\}_{i,j}),\bm{p}\tc \equiv \frac{2\pi^{\frac{d-J}{2}}}{\Gamma\ta\frac{d-J}{2}\tc}\int \prod_{i=1}^J dp_\para^i \int_0^\infty dp_\perp p_\perp^{d-J-1}f(p^2, \{\bm{p}\cdot \bm{q}_i\}_i,\{\bm{q}_i\cdot\bm{q}_j\}_{i,j}).
\]
Such a definition is consistent with the fundamental axioms of the map $\Id$ and justifies the commonly adopted nomenclature of `integration in $d$ dimensions'. 
Furthermore, it works independently on the dimension of $V$ so long as dim$(V)<+\infty$. 
One can easily extend the definition to the case of tensor integrals.
For a generic tensor function, one can always write down the following expansion into scalar functions $f_i(\bm{p}^2,\bm{q}^2,\bm{p}\cdot\bm{q})$,
\begin{align}
f^{ij}(\bm{p},\bm{q}) & = p^iq^j f_1(\bm{p}^2,\bm{q}^2,\bm{p}\cdot\bm{q})+q^ip^j f_2(\bm{p}^2,\bm{q}^2,\bm{p}\cdot\bm{q}) \notag \\
&+ p^ip^j f_3(\bm{p}^2,\bm{q}^2,\bm{p}\cdot\bm{q}) + q^iq^j f_4(\bm{p}^2,\bm{q}^2,\bm{p}\cdot\bm{q}) + g^{ij}f_5(\bm{p}^2,\bm{q}^2,\bm{p}\cdot\bm{q}),\label{TensorFuncDecomp}
\end{align}
and then proceed with the integration term by term.
If the index of the tensor function is carried by one of the external momenta $\bm{q}_i$, then there is no ambiguity in the meaning of the index, since the external momenta live in the finite subspace $V$. Vice versa, if the index is carried by $\bm{p}$, one enlarges the space $V$ for the integration of the corresponding term in Eq.~\eqref{TensorFuncDecomp}  in such a way to include that explicit component of $\bm{p}$ in the parallel integration. 
As final result, one has that in CDR or HV the open indices either belong to the minimal subspace $V$ which contains all the external vectors --- in such a case we usually have the identification between $V$ and the Wick-rotated Minkowski space --- or to the metric tensor $g^{\mu\nu}$.

The final step is to introduce a proper definition of the covariant tensor $g_{\mu\nu}$  in an infinite-dimensional space. The na\"ive construction as the inverse of $g^{\mu\nu}$ would lead to $g_{\mu\nu}g^{\mu\nu} = +\infty$, which is not very useful. Instead, one can define $g_{\mu\nu}$, and hence the dual space of covariant tensor operators, through the map $\Id$, by requiring that its action 
 on the generic tensor function $\bm{T}$ is
\[
g_{\mu\nu}(\bm{T}) = \frac{d\,\Gamma\!\ta\frac{d}{2}\tc}{\pi^{\frac{d}{2}}} \,\Id\!\ta \delta(\bm{p}^2-1)\sum_{i,j}T^{ij}p^ip^j,\bm{p}\tc.
\label{MetricDefinition}
\]
For the special case $\bm{T}=g^{\mu\nu}$ one obtains $g_{\mu\nu}g^{\mu\nu}=d$ as one would expect in a `$d$-dimensional space'. 

Within the framework of any DR scheme discussed above, the standard meaning of a component for any vector or tensor fails, notably if the component index is carried by the metric tensor which inherently lives in the infinitely-dimensional space QS$_d$.  However, this does not cause any trouble, since all physical observables are Lorentz scalars.
The component of a physical quantity represented by a vector or a tensor comes into play only through scalar products and assumes a particular value once  a reference frame has been specified. With this understanding, we have a clear interpretation  also in the DR schemes. 
\\[0.5cm]


\end{document}